\documentclass[acmsmall,screen]{acmart}

\setcopyright{rightsretained}
\acmPrice{}
\acmDOI{10.1145/3371117}
\acmYear{2020}
\copyrightyear{2020}
\acmJournal{PACMPL}
\acmVolume{4}
\acmNumber{POPL}
\acmArticle{49}
\acmMonth{1}

\usepackage{booktabs}   
\usepackage{subcaption} 

\usepackage{algorithm}
\usepackage[noend]{algpseudocode}
\usepackage{xspace}
\usepackage{amsmath}
\usepackage{tabularx,booktabs} 

\newtheorem{example}{Example}
\newtheorem{property}{Property}

\usepackage{listings}
\usepackage{subcaption}

\usepackage{stackengine}
\usepackage{bm}

\usepackage{bussproofs}
\EnableBpAbbreviations

\usepackage[inline]{aplcomments}
\newcommenter{chenglong}{1.0,0.4,1.0} 
\newcommenter{ras}{0.4,0.4,1.0} 
\newcommenter{yu}{1.0,0.4,0.4} 
\newcommenter{isil}{0.7,0.8,0.4} 
\newcommenter{alvin}{0.4,0.8,0.4} 

\definecolor{forestgreen}{RGB}{64,139,64}
\lstdefinestyle{falx}{ 
basicstyle=\footnotesize\ttfamily, 
breakatwhitespace=false, 
breaklines=true, 
captionpos=b, 
commentstyle=\itshape\color[rgb]{0.4,0.4,0.4}, 
deletekeywords={VALUE, INPUT}, 
firstnumber=1, 
frame=none, 
frameround=tttt, 
keywordstyle=\color{blue},
morekeywords={}, 
numbers=left, 
numbersep=3pt, 
numberstyle=\tiny\color[rgb]{0.5,0.5,0.5}, 
rulecolor=\color{black}, 
showstringspaces=false, 
showtabs=false, 
stepnumber=0, 
tabsize=2, 
backgroundcolor=\color{white},
}

\begin{document}

\title{Visualization by Example}         




\author{Chenglong Wang}
\affiliation{
  \institution{University of Washington}
  \country{USA}
}
\email{clwang@cs.washington.edu}          

\author{Yu Feng}
\affiliation{
  \institution{University of California, Santa Barbara}
  \country{USA}
}
\email{yufeng@cs.ucsb.edu}          

\author{Rastislav Bodik}
\affiliation{
  \institution{University of Washington}
  \country{USA}
}
\email{bodik@cs.washington.edu}          

\author{Alvin Cheung}
\affiliation{
  \institution{University of California, Berkeley}
  \country{USA}
}
\email{akcheung@cs.berkeley.edu}          

\author{Isil Dillig}
\affiliation{
  \institution{University of Texas at Austin}
  \country{USA}
}
\email{isil@cs.utexas.edu}          

\newcommand{\toolname}{\textsc{Viser}\xspace} 

\newcommand{\todo}[1]{{\color{red}{\{\{TODO: #1\}\}}}}
\newcommand{\chart}{\mathcal{V}}
\newcommand{\enc}{\textit{b}}
\newcommand{\tab}{\mathsf{T}}
\newcommand{\viz}{\textsf{V}}
\newcommand{\trace}{\tau}
\newcommand{\spec}{\textsf{P}}
\newcommand{\inexample}{\mathcal{I}}
\newcommand{\outexample}{\mathcal{O}}
\newcommand{\eval}[1]{{[\![#1]\!]}}
\newcommand{\peval}[1]{{[\![#1]\!]_\partial}}
\newcommand{\pevalinv}[1]{{[\![#1]\!]^{-1}_\partial}}
\newcommand{\constraint}{\phi}
\newcommand{\comp}{\mathcal{X}}
\newcommand{\comps}{\Gamma}
\newcommand{\prog}{\mathsf{P}}
\newcommand{\preds}{\Sigma} 
\newcommand{\abs}{\alpha}
\newcommand{\concate}{\mathbin{+\mkern-10mu+}} 
\newcommand{\lang}{\mathcal{L}}
\newcommand{\symbolic}[1]{{\tilde{#1}}}
\newcommand{\veval}[1]{{[\![#1]\!]_\viz}}
\newcommand{\vinverse}[1]{{[\![#1]\!]_{\viz}^{-1}}}
\newcommand{\sample}{\mathcal{E}}
\newcommand{\code}[1]{\texttt{\small #1}}
\newcommand{\hole}{\square}
\newcommand{\textblue}[1]{{\textcolor{blue}{#1}}}
\newcommand{\substcontain}{\mathrel{\ensurestackMath{\stackon[1pt]{\subseteq}{\scriptstyle\diamond}}}}

\newcommand{\vscript}{\textsf{P}}
\newcommand{\pvis}{\mathsf{P}_\mathsf{V}}
\newcommand{\ptbl}{\mathsf{P}_\mathsf{T}}
\newcommand{\tbl}{{\textsf{T}}}
\newcommand{\atbl}{\tilde{\tbl}}
\newcommand{\intbl}{\tbl_{\textsf{in}}}
\newcommand{\vsketch}{\mathsf{S}}
\newcommand{\dsource}{\mathsf{I}}
\newcommand{\outtbl}{\tbl_{\textsf{out}}}
\newcommand{\mult}{\textsf{Mult}}
\newcommand{\myvec}[1]{\bar{#1}}
\newcommand{\tblconstraint}{\psi}
\newcommand{\partialprog}{\mathbb{P}}
\newcommand{\forward}{\downarrow}
\newcommand{\backward}{\uparrow}
\newcommand{\inconstraint}{\tblconstraint_{\mathsf{in}}}
\newcommand{\outconstraint}{\tblconstraint_{\mathsf{out}}}
\newcommand{\tblincl}{\phi}
\newcommand{\toolvar}{\toolname{\sc M}}

\newcommand{\tsketch}{\psi_{\tab_\text{out}}}
\newcommand\concat{+\kern-1.3ex+\kern0.8ex}

\newsavebox{\fmbox}
\newenvironment{smpage}[1]
{\begin{lrbox}{\fmbox}\begin{minipage}{#1}}
{\end{minipage}\end{lrbox}\usebox{\fmbox}}
\newcommand{\worklist}{W}
\newcommand{\semantic}{\Psi}

\begin{abstract}
While  visualizations play a crucial role in gaining insights from data, 
generating useful visualizations from a complex dataset is far from an easy task. In particular, besides
understanding the functionality provided by existing visualization libraries, 
generating the desired visualization also requires reshaping and aggregating the 
underlying data as well as composing different visual elements to achieve 
the intended visual narrative. This paper aims to simplify visualization tasks
by automatically synthesizing the required program from simple \emph{visual sketches}
provided by the user. Specifically, given an input data set and a visual sketch that 
demonstrates how to visualize a very small subset of this data, our technique automatically
generates a program that can be used to visualize the entire data set.

From a program synthesis perspective, automating visualization tasks poses several challenges
that are not addressed by prior techniques. First, because many visualization tasks 
require data wrangling in addition to generating plots from a given table, we need to 
 decompose the end-to-end synthesis task into two separate sub-problems. Second, because
 the intermediate specification that results from the decomposition is necessarily imprecise, this makes
 the data wrangling task particularly challenging in our context. In this paper, we address these problems by
 developing a new \emph{compositional} visualization-by-example technique that (a) decomposes
 the end-to-end task into two different synthesis problems over different DSLs and (b) leverages 
 bi-directional program analysis to deal with the complexity that arises 
 from having an imprecise intermediate
 specification. 
 
 We have implemented our visualization-by-example approach in a tool called \toolname\ and evaluate
 it on 83 visualization tasks collected from on-line forums and tutorials. \toolname can solve 
$84\%$ of these benchmarks within a 600 second time limit, and, for those tasks that can be solved, 
 the desired visualization is among the top-$5$ generated by \toolname in $70\%$ of the cases.

\end{abstract}



\maketitle

\section{Introduction}\label{sec:intro}

Visualizations play an important role in today's data-centric world for discovering, validating,  and communicating insights from data. 
Due to the prevalence of non-trivial visualization tasks across different application domains, recent years have seen a growing number of  libraries that aim to facilitate complex visualization tasks. For instance, there are at least a dozen different visualization libraries for Python and R, and more than ten different visualization libraries for JavaScript have emerged in the past year alone~\footnote{\url{https://financesonline.com/data-visualization/}}. In addition, there has also been a flurry of research activity around building programming systems like D3~\cite{DBLP:journals/tvcg/BostockOH11} and Vega-Lite~\cite{DBLP:journals/tvcg/SatyanarayanMWH17} to further facilitate real-world visualization tasks.

Despite all these recent efforts, data visualization still remains a challenging task that requires considerable expertise --- in fact, so much so that some companies even  have job titles  like  ``data visualization expert'' or ``data visualization specialist.'' Generally speaking, there are three key reasons that make data visualization a challenging task. First, beyond having a good insight about how the data can be best visualized, one needs to have sufficient knowledge about how to use the relevant visualization libraries. Second, different visualization primitives typically require the data to be in different formats; so, in order to experiment with different types of visualizations, one needs to constantly reshape the data into different formats. Finally, generating the intended visualization typically requires modifications to the original dataset, including aggregating and mutating values and adding new columns to the input tables, and doing so often require deep knowledge in data manipulation.

In this paper, we propose a new technique, coined \emph{visualization-by-example}, for automating data visualization tasks using program synthesis. In our proposed approach, the user starts by providing a so-called \emph{visual sketch}, which is a partial visualization of the input data for just a few input points.
Given the original data set $\intbl$ and a visual sketch $\vsketch$ provided by the user, our technique can synthesize one or more visualization scripts whose output is consistent with $\vsketch$ for the input data set. These visualization scripts can then be applied to $\intbl$ to generate several visualizations of the \emph{entire data set}, and the user can choose the desired visualization among the ones that are generated. 

Despite these appealing aspects of our approach to end-users, the data visualization problem presents unique challenges from a program synthesis perspective. First, as hinted earlier, data visualization tasks almost always involve two distinct steps, namely (1) data wrangling (reshaping, aggregating, adding new columns etc.) and (2) invoking the appropriate  visualization primitives on the transformed data. Asking users to manually decompose the problem into these two individual steps would  defeat the point,
as  the user  would have to at least understand which visualization primitives to use and what format they require. Thus, it is imperative to have a compositional technique that can \emph{automatically decompose} the end-to-end task into two separate synthesis problems.

One of the key contributions of this paper is to show how to automatically decompose an end-to-end visualization task into two separate synthesis problems over two different languages. Specifically, given an input data source $\intbl$ and a visual sketch $\vsketch$, our goal is to learn a \emph{table transformation program} $\ptbl$ and a \emph{visual program} $\pvis$ such that executing $\pvis \circ \ptbl$ on $\intbl$ yields a visualization that is consistent with the provided visual sketch $\vsketch$. In order to solve this problem in a compositional way, our method infers an intermediate specification $\phi$ that constrains the output (resp. input) of the target program $\ptbl$ (resp. $\pvis$). This intermediate specification is in the form of \emph{table inclusion constraints} $\tbl \substcontain t$ specifying that  input  $t$ of the visual program must include all tuples in $\tbl$ but it can also contain additional rows and columns. As we demonstrate experimentally, having this intermediate specification is crucial for the scalability of our approach. 

A second key contribution of this paper is a new algorithm for synthesizing table transformation programs. While there has been recent work on automating table transformation tasks using  programming-by-example~\cite{scythe,morpheus}, these techniques focus on the case where  the specification is a pair of input and output tables. In contrast, the intermediate specification  in our setting is a set of table inclusion constraints  rather than a concrete output table, and pruning strategies used in prior work are not effective in this setting due to lack of precise information about the output table. To deal with this challenge, we introduce a new table transformation synthesis algorithm that uses lightweight bidirectional program analysis to prune the search space. As we demonstrate experimentally, this new table transformation algorithm results in much faster synthesis compared to prior work~\cite{morpheus} for automating visualization tasks.

We have implemented the proposed ``visualization-by-example'' approach in a new tool called \toolname and evaluated it on 83 visualization tasks collected from on-line forums and tutorials. Our experiments show that \toolname can solve $84\%$ of these benchmarks and, among those benchmarks that can be solved,  the desired visualization is among the first $5$ outputs generated by \toolname in $70\%$ of the cases. Furthermore, given that it takes on average of 11 seconds to generate the top-5 visualizations, we believe that \toolname is fast enough to be beneficial to prospective users in practice.

To summarize, this paper makes the following key contributions:

\begin{itemize}
\item We introduce the \emph{visualization-by-example} problem and present an algorithm for synthesizing visualization scripts given the original data set  and a small visual sketch.
\item We show how to decompose the synthesis task into two sub-problems by inferring an intermediate specification in the form of table inclusion constraints.
\item We propose a new algorithm for synthesizing table transformations from table inclusion specifications. Our algorithm leverages lightweight bidirectional program analysis to effectively prune the search space.
\item We  evaluate our approach on over 80 tasks collected from on-line forums and tutorials and show that  \toolname can solve $84\%$ of the benchmarks, and that the desired visualization is among the top-$5$ results in $70\%$ of the solved cases. 
\end{itemize}

\section{Overview}\label{sec:overview}
\begin{figure}[t]
  \centering
  \includegraphics[width=\textwidth]{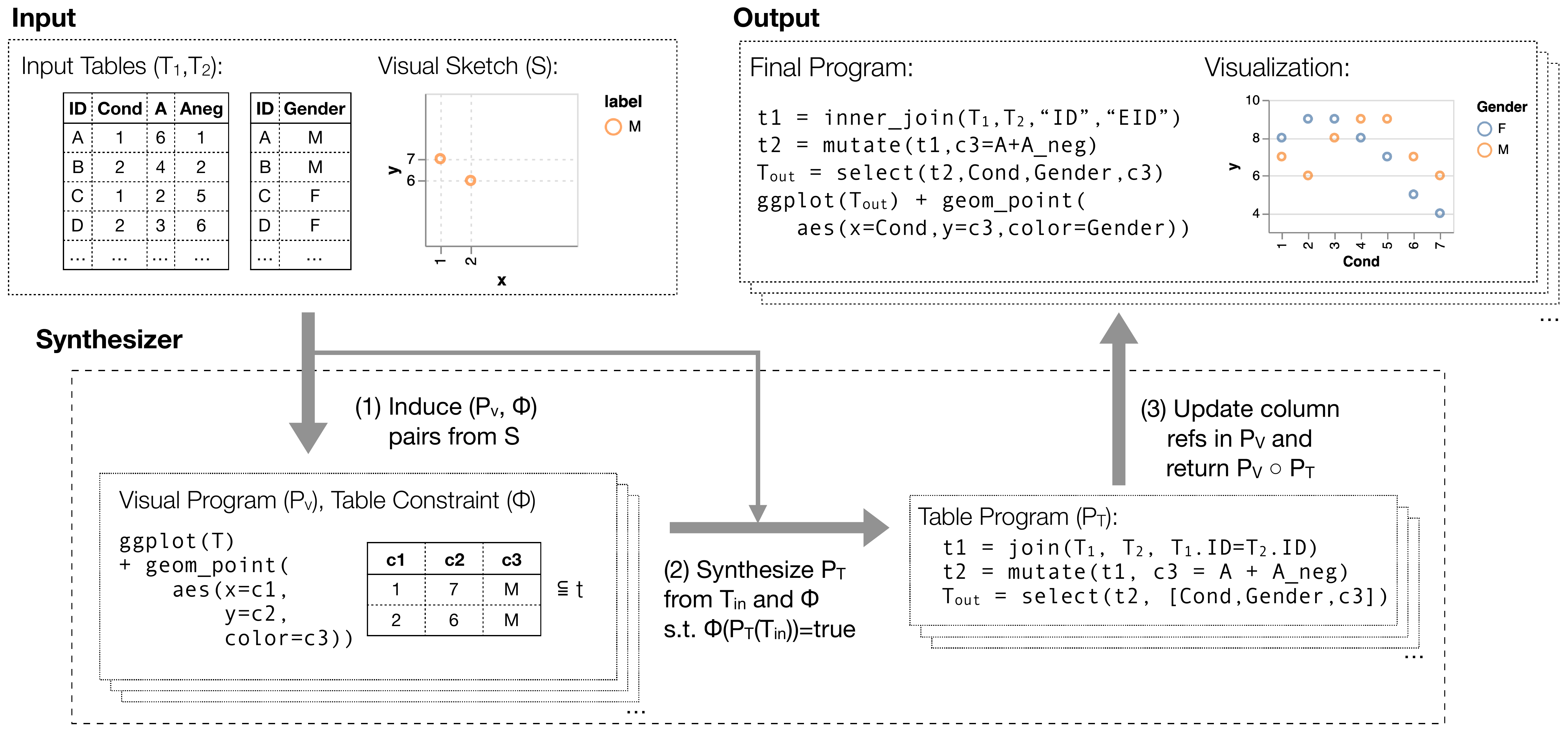}
\caption{Overview of our synthesis algorithm: the system takes as input an input table $\tab_\mathsf{in}$ and a visual sketch $\vsketch$, and returns a list of candidate visualizations satisfying the inputs.}
\label{fig:overview}
\end{figure}

In this section, we give an overview of our approach with the aid of a simple motivating example depicted in \autoref{fig:overview}. In this  example, the user has two tables $\tab_1$ and $\tab_2$ that record the results of a scientific experiment. Specifically,  Table $\tab_1$ stores an experiment identifier (\code{ID}), a so-called "experiment condition" (\code{Cond}), and the experiment result, which consists of an  \code{A}  value as well as an  \code{Aneg} value.  An additional table $\tab_2$ stores the gender of the participant in the corresponding study: That is, for each experiment (\code{ID}), the \code{Gender} column in $\tab_2$ indicates  whether the participant is male (M) or female (F).   The user wants  to visualize the result of this experiment by drawing a scatter plot that shows how the sum of \code{A} and \code{Aneg} changes with respect to \code{Cond} for each of the two genders. In particular, the top right part of \autoref{fig:overview} illustrates the desired visualization. In the remainder of this section, we explain how our approach synthesizes the desired visualization script in R using the {\tt tidyverse} package collection, which includes both visualization primitives (e.g., provided by {\tt ggplot2}) and data wrangling capabilities (e.g., provided by {\tt tidyr} and {\tt dplyr}).

\paragraph{User input} 
In order to use our visualization tool, \toolname, the user needs to provide the data source  (i.e., tables $\tab_1$ and $\tab_2$) as well as a visual sketch $\vsketch$, which is a partial visualization for a tiny subset of the original data source.  In this case, since the user wants to draw a scatter plot, the visual sketch is very simple and consists of a few data points, as shown in the "Input" portion of \autoref{fig:overview}. Specifically, the visual sketch contains two points, one at  $(1, 7)$ and the other at $(2,6)$, and both points have label "\code{M}". In general, one can think of a visual sketch as a set of \emph{visualization elements} (e.g., point, bar, line, \ldots) where each visualization element has various attributes, such  as coordinates, color, etc. In particular, we can specify the visual sketch shown in \autoref{fig:overview} as the following set of visual elements:
$$\{ \mathsf{point}(v_{x}=1, v_{y}=7, v_\mathit{color}=\code{M}),
                      \mathsf{point}(v_{x}=2, v_{y}=6, v_\mathit{color}=\code{M}) \} $$
We refer to this alternative set-based representation of a visualization as a \emph{visual trace}. In general, we can represent both visual sketches as well as complete visualizations in terms of their corresponding visual trace; thus, we use the term "visual trace" interchangeably with both "visualization" and "visual sketch".

\paragraph{Synthesis problem and approach} Given the input data source $\dsource$ and a visual sketch $\vsketch$, our synthesis problem is to infer a \emph{visualization script} $\vscript$ such that $\vscript(\dsource)$ yields a visual trace that is \emph{a superset} of $\vsketch$. As mentioned in \autoref{sec:intro}, a visualization script consists of a pair of programs $\pvis$ (``visual program'') and $\ptbl$ (``table transformation program''), for plotting and data wrangling respectively. Since $\pvis$ and $\ptbl$ do conceptually different things and are expressed in separate languages, we decompose the overall synthesis task into two sub-tasks, namely that of synthesizing a visual program $\pvis$ and separately synthesizing a table transformation program $\ptbl$.

\begin{figure}[t]
  \centering
  \includegraphics[width=\textwidth]{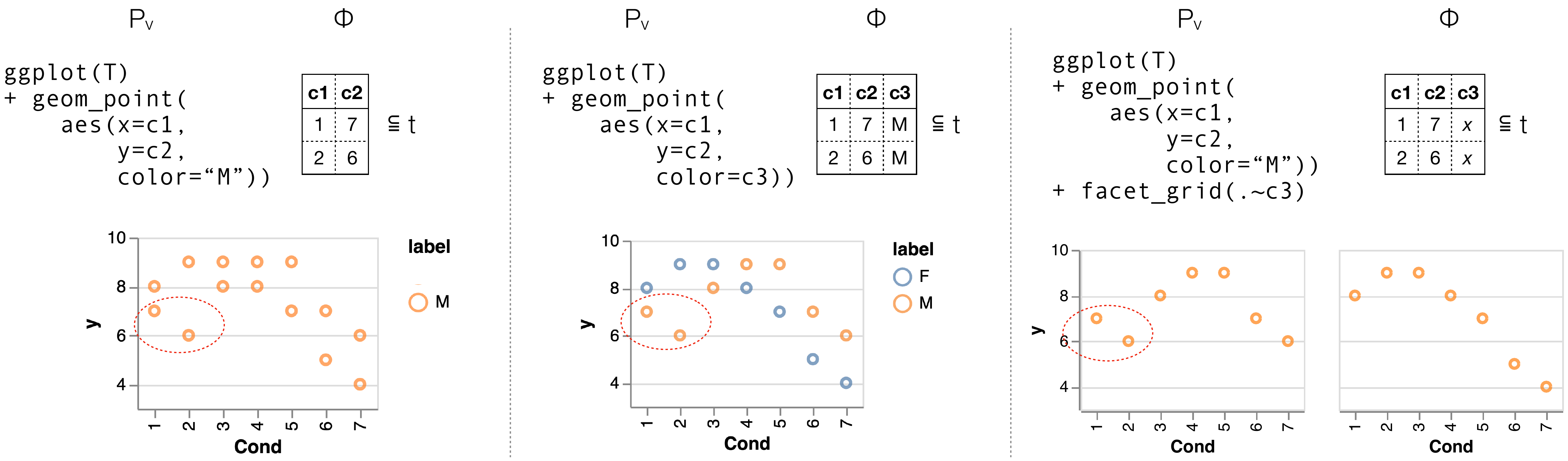}
\caption{Sample visual programs and their corresponding intermediate specification for the visual sketch from  \autoref{fig:overview}. The bottom part of each figure shows the corresponding visualization if the visual program is adopted; observe that all of these visualizations are consistent with the visual sketch.} 
\label{fig:vis-inv-example}
\end{figure}

\paragraph{Synthesis of visual programs.}
To achieve the decomposition outlined above, our synthesis algorithm first infers a \emph{set} of visual programs that are \emph{capable} of generating a visualization consistent with $\vsketch$. This inference step is based purely on the visual sketch and does not consider the input data (i.e., tables $\tab_1$, $\tab_2$).  For our running example, there are multiple visual programs (expressible using \code{ggplot2}) that can generate the desired result; we show three of these programs in \autoref{fig:vis-inv-example}. All three programs start with the code ``\code{ggplot(T) + \code{geom\_point(...)}}'', which indicates that the resulting visualization is a scatter plot drawn from  table $T$. However, the three visual programs differ in the following ways:

\begin{itemize}
\item All points generated by the first visual program have the same color (indicated as \code{color="M"})
\item For the second visual program, the color of the points is determined by the corresponding value of column \code{c3} in table \code{T} (indicated as \code{color=c3})
\item The visualization generated by the last  program contains multiple subplots determined by $c_3$. That is,  the visualization is partitioned into a list of subplots according to different values in column $c_3$ of the input table.
\end{itemize}
As indicated in the bottom part of \autoref{fig:vis-inv-example}, the visualizations generated by all three programs are consistent with the visual sketch in that they contain the two data points specified by the user.

\paragraph{Intermediate specification inference.}
 Next, given the visual sketch $\vsketch$ and a  candidate visual program   $\pvis$, our synthesis algorithm infers an intermediate specification  $\tblincl$ that constrains the input that $\pvis$ operates on. Furthermore, $\tblincl$ has the property that executing $\pvis$ on any concrete table consistent with $\intbl$ yields a visual trace that is a superset of $\vsketch$. 
Going back to our running example, \autoref{fig:vis-inv-example} shows the intermediate specifications inferred for each of the three visual programs. These intermediate specifications are of the form $\tbl \substcontain t$ indicating that the input $t$ of the visual program must contain table $\tbl$ but can also include additional rows and columns. Looking at the intermediate specifications from \autoref{fig:vis-inv-example}, we can make the following observations:

\begin{itemize}
\item The inputs of the first visual programs must contain at least two columns (referred to as \code{c1}, \code{c2}) and these columns should contain the values 1, 2 and 7, 6 respectively. However, the input table can also contain additional attributes and values.
\item The specification for the second visual program imposes one additional constraint over the other ones. In particular, the input table must contain an  additional third column (referred to as \code{c3}), and this column  must contain at least two occurrences of value \code{M}.
\item The specification for the third visual program requires that the table should contain an additional column \code{c3} to specify which subplot  each point belongs to. Since the visual sketch contains two points in the same subplot, column \code{c3} contains two duplicate values with the same subplot identifier.
\end{itemize}

\begin{figure}[t]
  \centering
  \includegraphics[width=\textwidth]{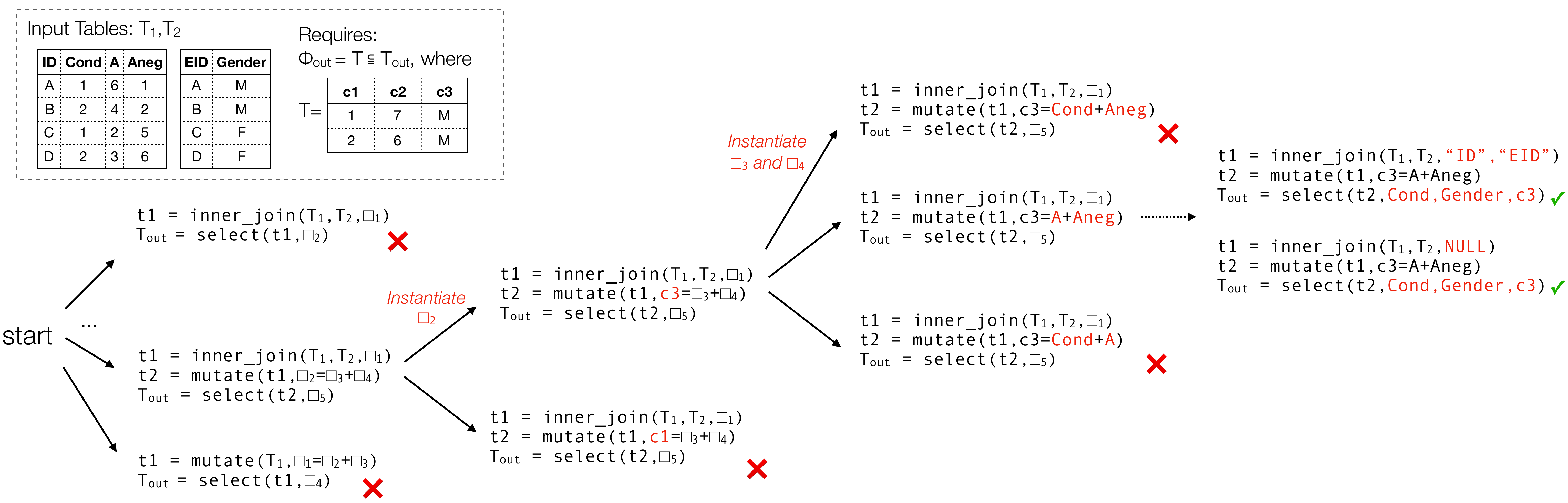}
\caption{The synthesis process for $\ptbl$ using $\tab_1,\tab_2$ and $\phi_\mathsf{out}$ in \autoref{fig:overview}. At each step, the synthesis algorithm first picks a known variable and expands it (new values expanded at each step are labeled in \textcolor{red}{red}), then it evaluates each program sketch using abstract semantics of the table transformation language and prune it if the evaluation process results in conflicts.}
\label{fig:alg-example}
\end{figure}

\paragraph{Input for the second synthesis task.} As mentioned earlier, the key reason for inferring an intermediate specification is to 
decompose the problem into two separate synthesis tasks. Thus, given an initial data source $\intbl$ and intermediate specification $\tblincl$, the goal of the second synthesis task is to generate a table transformation program $\ptbl$ such that applying $\ptbl$ to $\intbl$ yields a table that is consistent with $\tblincl$. 
To illustrate how our method synthesizes the desired table transformation program for our running example, let us consider the following data wrangling constructs:

\begin{itemize}
\item {\bf Projection:} The construct $\code{select}(t, \bar{c})$ computes the projection of table $t$ onto columns  $\bar{c}$.
\item{\bf Join:}  The construct $\code{inner\_join}(t_1, t_2, p)$ computes the product of  tables $t_1, t_2$ and then filters the  result based on predicate $p$.
\item {\bf Mutation:} The construct $\code{mutate}(t, c_\mathit{target}, c_\mathit{arg_1} + c_\mathit{arg_2})$ takes as input a table $t$ and returns a table with a new column called $c_\mathit{target}$, where the values in $c_\mathit{target}$ are obtained by summing up the two columns $c_\mathit{arg_1}$ and $c_\mathit{arg_2}$;~\footnote{General {\tt mutate} operator supports arbitrary column-wise computation besides `+', we only consider {\tt mutate} with `+' in overview for simplicity.}
\end{itemize}

We will now illustrate how to synthesize the desired table transformation program $\ptbl$ for the input tables $\tbl_1, \tbl_2$ shown in \autoref{fig:overview} and the intermediate specification $\tblincl_\mathsf{out}$ shown in the second column of \autoref{fig:alg-example}.

\paragraph{Table transformation synthesis overview.} {\toolname} employs an enumerative search algorithm to find a table transformation program that satisfies the specification. Similar to prior program synthesis techniques~\cite{morpheus,scythe,DBLP:conf/pldi/FengMBD18}, \toolname uses lightweight deductive reasoning  to prune invalid programs during the search process. However, because the specification does not involve a concrete output table, pruning techniques used in prior work (e.g., \cite{morpheus}) are not effective in this context.  As mentioned in \autoref{sec:intro}, our algorithm addresses this issue by leveraging lightweight bidirectional program analysis and an (also lightweight) incomplete inference procedure over table inclusion constraints. 

As illustrated schematically in \autoref{fig:alg-example}, \toolname performs enumerative search over \emph{program sketches}, where each program sketch is a sequence of statements of the form \code{v = op($\Box_1, \ldots, \Box_n$)} where \code{op} is one of the data wrangling constructs (e.g., \code{select}, \code{inner\_join} etc.) and $\Box_i$ denotes an unknown argument.  Since  program sketches contain only table-level operators but not their arguments, the sketch enumeration process is tractable, and the main synthesis burden lies in searching the large number of parameters that each hole $\Box_i$ can be instantiated with.

\paragraph{Sketch completion} Given a program sketch, \toolname's sketch completion procedure alternates between \emph{hole instantiation} and \emph{pruning} steps until a solution is found or all possible sketch completions are proven \emph{not} to satisfy the specification (see \autoref{fig:alg-example}). The first step (i.e., hole instantiation) is standard and makes the initial sketch iteratively more concrete by filling each hole with a program variable or constant. The pruning step, on the other hand, is more interesting, and  infers {table inclusion constraints} of the form $e_1 \substcontain e_2$ indicating that the table represented by expression $e_1$ can be obtained from the table represented by expression $e_2$ by removing rows and/or columns. For the forward (resp. backward) inference, the generated constraints have the shape $t \substcontain \tbl$ (resp. $\tbl \substcontain t$), where $t$ is a program variable and $\tbl$ is a concrete table. Thus, using a combination of forward and backward reasoning, we can obtain "inequality" constraints of the form $\tbl_1 \substcontain t \substcontain \tbl_2$ for each program variable $t$ and use this information to reject a \emph{partially completely sketch}  whenever $\tbl_1$ \emph{cannot} be obtained from $\tbl_2$ by deleting rows and/or columns.

\begin{figure}[t]
  \centering
  \includegraphics[width=1\textwidth]{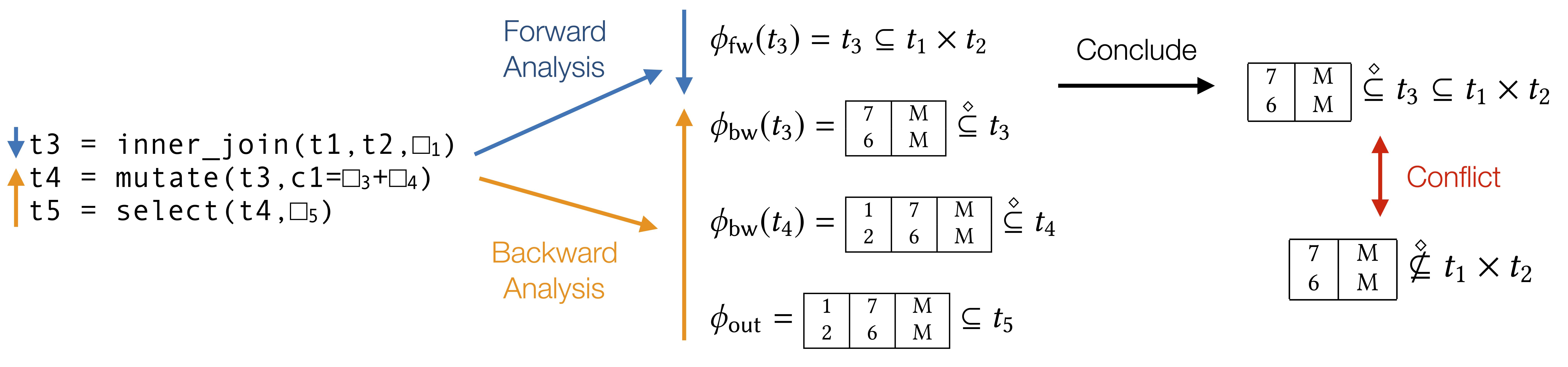}
\caption{Demonstration of how \toolname prunes invalid abstract programs in \autoref{fig:alg-example} using forward and backward analysis. $\phi_\mathsf{out}$ is the requirement from the synthesis task, and $\phi_\mathsf{fw}(t), \phi_\mathsf{bw}(t)$ refers to abstractions of $t$ derived from forward / backward analysis.}
\label{fig:alg-pruning-example}
\end{figure}

Going back to our running example, let us consider the partially completed sketch shown in \autoref{fig:alg-pruning-example}, where $t_1, t_2$ represent the program's arguments and $t_5$ is the return variable.  By considering the semantics of each construct, we can make the following deductions:

\begin{enumerate}\itemsep3pt
\item Since the program's output must conform to  $\tblincl_\mathsf{out} = (\tbl \substcontain t)$ (where $\tbl$ is the table from  \autoref{fig:vis-inv-example}) and we have $t=t_5$, we can generate the constraint $\tbl \substcontain t_5$ on variable $t_5$.
\item Together with the above constraint and semantics of \code{select}, we obtain   $\tbl \substcontain t_4$.
\item Since we have $\tbl \substcontain t_4$ and \code{mutate(t3, c1, ? + ?)} generates $t_4$ by adding column \code{c1} to $t_3$, we can deduce all columns of $\tbl$ except for the first one should also be in $t_3$. Thus, using backwards reasoning, we obtain the constraint $\tbl' \substcontain t_3$ where $\tbl'$ is the table shown on the left-hand side of $\phi_\mathsf{bw}(t_3)$.
\item  Since $t_3$ is the result of \code{inner\_join(t1, t2, ?)}, we cannot deduce something useful about $t_1, t_2$ in the backward direction without introducing expensive case splits because we do not whether  each value in $\tab'$ comes from $t_1$ or $t_2$.  However, going in the \emph{forward} direction,  we can generate the constraint $t_3 \substcontain T_1 \times T_2$, where $T_1, T_2$ are the input tables from \autoref{fig:overview}.
\end{enumerate}

Putting together the information obtained from the forwards and backwards analysis, we obtain the constraint $\tbl' \substcontain t_3 \substcontain \tbl_1 \times \tbl_2$. However, this creates a contradiction with  $\tbl' \not \substcontain \tbl_1 \times \tbl_2$, allowing us to prune the partially completed sketch from \autoref{fig:alg-pruning-example}.

\paragraph{Synthesis output.} Continuing in this manner and alternating between more hole instantiation and pruning steps, \toolname finds the sketch completion shown on the bottom right  side of \autoref{fig:overview}.  The final output of the synthesizer is shown on the top right of the same figure and generates the intended visualization, also shown in the top right of \autoref{fig:overview}.

\section{Problem Definition}\label{sec:problem}
In this section, we formally define the visualization-by-example problem and then introduce two languages for automating table transformation and plotting tasks.

\subsection{Key Concepts and Synthesis Problem}~\label{sec:prob-def}

\paragraph{\bf \emph{Tables}}  For the purposes of this paper, a table $\tab$ with schema $[c_1,\dots,c_n]$  is an unordered bag (i.e., multi-set) of tuples where each tuple $r=(v_1,\dots,v_n)$ in $\tab$ consists of $n$ primitive values (number, string,  datetime etc.).
Given a tuple $r\in \tab$, we use the notation $r[c]$ to denote the value $v$ stored in attribute $c$ of $r$. We also extend this notation to tables and write  $T[\myvec{c}]$ to denote the projection of table $\tab$ on columns $\myvec{c}$ and  write  $T[-\myvec{c}]$ to denote the projection of table $\tab$ on all columns \emph{except} $\myvec{c}$.  Finally, we write $\mult(r, \tab)$ to denote the multiplicity of tuple $r$ in $\tab$. 

Given a pair of tables $\tab_1, \tab_2$, we write  $\tab_1\subseteq\tab_2$  iff $\forall r\in \tab_1. \ \mult(r, \tab_1) \leq \mult(r, \tab_2)$. As standard, we define equality to be containment in both directions, i.e.,  $\tab_1=\tab_2$ iff $\tab_1\subseteq \tab_2$ and $\tab_2\subseteq \tab_1$. We further define a table inclusion constraint $\tab_1\substcontain \tab_2$ that allows projecting columns in addition to filtering rows. Specifically, we write $\tab_1\substcontain\tab_2$ iff there exists columns $\myvec{c}$ in the schema of $\tab_2$ such that $\tab_1 \subseteq \tab_2[\myvec{c}]$.

\begin{figure}[ht]
\[
\begin{array}{rlll}
\mathit{\trace} & = & \{\mathit{e}_1, \dots, \mathit{e}_n\}\\
\mathit{e} & = 
       & \mathsf{bar}(a_{x}, a_{y_1}, a_{y_2}, a_\mathit{color}, a_\mathit{subplot})\\
      &|& \mathsf{point}( a_{x}, a_{y}, a_\mathit{color}, a_\mathit{size}, a_\mathit{subplot})\\
      &|& \mathsf{line}(a_{x_1}, a_{y_1}, a_{x_2}, a_{y_2}, a_\mathit{color}, a_\mathit{subplot})
\end{array}
\]
\vspace{-10pt}
\caption{The visualization trace language $\lang_{\trace}$, where metavariable $a$ refers to constants.}
\label{fig:trace-lang}
\end{figure}

\paragraph{\bf \emph{Visual traces}} As stated in Section~\ref{sec:overview}, we define the semantics of visualizations in terms of so-called \emph{visual traces}. A visual trace, denoted $\trace$, is a set of basic visual elements ($\mathsf{point}$, $\mathsf{line}$, $\mathsf{bar}$), together with the  attributes of each element (position, size, color, etc.). More concretely, \autoref{fig:trace-lang} shows a small ``language'' in which we express visual traces. (The full language supported by our implementation is given in the Appendix.)  Here, $e$ denotes a visual element, and $a$ is an \emph{attribute} of that element:

\begin{itemize}
\item \emph{Color attribute:} This attribute, denoted $a_\emph{color}$ specifies the color of a visual element.
\item \emph{Position attributes:}  Position attributes, such as $a_x, a_{x_1}, a_{y_2}$ etc., specify the canvas positions for a visual element. For example, for \textsf{line}, $(a_{x_1}, a_{y_1})$ specifies the starting point of a line segment, and  $(a_{x_2}, a_{y_2})$ specifies the end point. For the \textsf{bar} visual element, $a_{y_1}, a_{y_2}$ specify the start and end $y$-coordinates of a (vertical) bar.
\item  \emph{Size attribute:} The attribute $a_\emph{size}$ specifies the size of a given \textsf{point} element.
\item \emph{Subplot attribute:} The attribute $a_\emph{subplot}$ specifies the subplot that a given visual element belongs to. For instance, for the visualization shown in the last column of \autoref{fig:vis-inv-example}, the points in the first plot have a different $a_\emph{subplot}$ attribute than those in the second one.
\end{itemize}

In the remainder of this paper, we express both complete visualizations and visual sketches in terms of their corresponding visual trace, and we often use the symbol $\vsketch$ to denote traces that correspond to visual sketches. Finally, since visual traces are \emph{sets} of visual elements, the notation $\trace_1 \subseteq \trace_2$ indicates that visualization $\trace_2$ is an \emph{extension} of visualization $\trace_1$.

\paragraph{\bf \emph{Problem statement}} Given this notion of visual traces, we can now state our \emph{visualization-by-example problem}, which is defined by a pair $(\intbl, \vsketch)$. Here,  $\intbl$ is a table~\footnote{We note that our implementation can handle multiple  tables in the input. However,  we consider a single input table in the formal development to simplify presentation and reduce notational overhead.}  and  $\vsketch$ is a visual trace (i.e., a "program" in the language of Figure~\ref{fig:trace-lang}). Now, let us fix  a language  $\lang_\tab$ for expressing table transformation programs and a visualization language $\lang_\viz$ for generating plots from a given table. Then, our goal is to synthesize a pair  of programs $(\ptbl, \pvis)$ such that:
\begin{enumerate}
\item $\ptbl$ and $\pvis$ are programs written in $\lang_\tab, \lang_\viz$ respectively,
\item the output visual trace $\ptbl(\pvis(\tab_\text{in}))$ is consistent with $\vsketch$, i.e., $\vsketch \subseteq \pvis(\ptbl(\tab_\text{in}))$.
\end{enumerate}

Note that our problem statement strictly generalizes  conventional programming-by-example which  requires the program output to be equal to the provided output (i.e.,  $\vsketch = \pvis(\ptbl(\tab_\text{in}))$). Thus, the user is still free to provide  a small (but complete) input-output example if this is more convenient for the user. However, our generalization has the advantage of freeing the user from the burden of modifying the input data in cases where doing so may be inconvenient.

\subsection{Table Transformation Language}\label{sec:table-dsl}
Our table transformation language  is shown in \autoref{fig:tbl-lang}. This language is inspired by existing data wrangling libraries (e.g., tidyr and dplyr libraries for R), and similar languages have also been used in prior work for automating table transformation tasks using PBE~\cite{morpheus,trinity}.  As shown in \autoref{fig:tbl-lang}, a table transformation program $\ptbl$ is a sequence of side-effect free statements, where each statement produces a new table by performing some operation on its inputs.  As standard in relational algebra, the constructs \textsf{select} and \textsf{filter} are used for selecting columns and rows respectively. As also standard in relational algebra, \textsf{join} is used for taking the cross product of two tables. That is, \textsf{join}($T_1, T_2, f$) is semantically equivalent to \textsf{filter}($T_1 \times T_2, f$), where $\times$ denotes the standard cross product operator in relational algebra.

\begin{figure}[!t]
\[\begin{array}{rlll}
  \ptbl(t) & = & t_1=e_1;\dots;\tab_\text{out}=e_n;\\
  e & = & T ~|~ \mathsf{filter}(T, f) ~|~ \mathsf{select}(T, \bar{c}) ~|~ \mathsf{join}(T_1, T_2, f)\\
    &|& \mathsf{mutate}(T, c_\mathit{target}, \mathit{op}, \bar{c}_\mathit{arg}) ~|~ \mathsf{gather}(T, \bar{c}_\mathit{id}, \bar{c}_\mathit{target}) \\
    &|& \mathsf{spread}(T, \bar{c}_\mathit{id}, c_\mathit{key}, c_\mathit{val}) \\
    &|& \mathsf{summarize}(T, \bar{c}_\mathit{key}, \alpha, c_\mathit{target})
\end{array}
~
\begin{array}{rlll}
  T & = & \tab_\text{in} ~|~ t\\
  f & = & v_1~\mathit{op}~v_2 ~|~ \mathsf{is\_null}(c)\\
  v & = & \mathsf{const} ~|~ c\\
  \alpha & = & \mathsf{min}~|~ \mathsf{max}\\
         & | & \mathsf{sum}~|~ \mathsf{count} ~|~ \mathsf{avg}
\end{array}
\]
\vspace{-10pt}
\caption{The table transformation language $\lang_{\tab}$, where $\tab_\text{in},\tab_\text{out}$ refers to the input/output tables, $t$ refers to table variables, and $c$ refers to column names.}
\label{fig:tbl-lang}
\end{figure}

\begin{figure}[!t]
\includegraphics[width=0.9\textwidth]{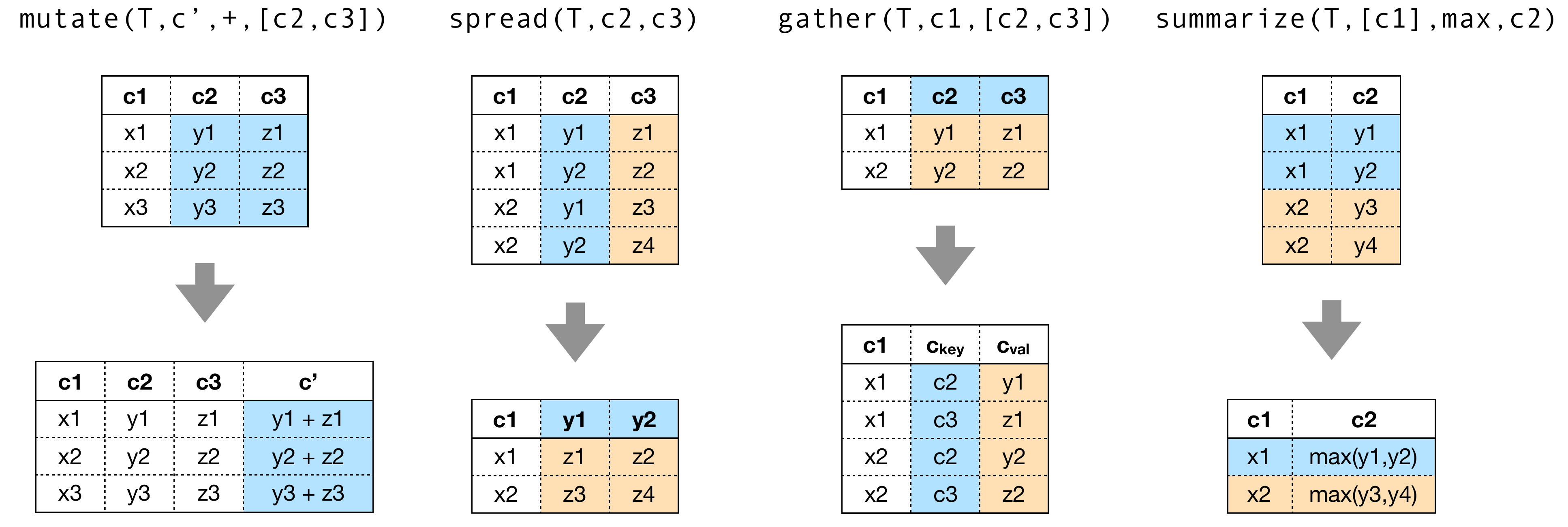}
\caption{Examples of table transformation operators in $\lang_\tab$ and how they operate on example input tables. Colored cells shows how parts of the output table are computed from the input table.}
\label{fig:ptbl-example}
\end{figure}

Besides these standard relational algebra operators, our table transformation language contains four other main constructs, namely \textsf{spread, mutate},  \textsf{gather}, and \textsf{summarize}. Since the semantics of these constructs are somewhat non-trivial, we illustrate their behavior in \autoref{fig:ptbl-example}. 

\begin{enumerate}\itemsep-1pt
\item The \code{mutate} construct creates a new column ($c_\mathit{target}$) in the output table by applying  an operator \emph{op} on argument columns $\bar{c}_\mathit{args}$. For example, in \autoref{fig:ptbl-example}, the new column $c'$ is obtained by  summing up columns $c_2$ and $c_3$.
\item The \code{spread} operator pivots a table by changing values  to column names. Specifically, \code{spread}  first eliminates the two columns $c_\mathit{key}$ and $c_\mathit{val}$, then creates a new column for 
each value stored in the original column $c_\mathit{key}$, and finally fills new columns using values in the original $c_\mathit{val}$ column.   The second drawing in \autoref{fig:ptbl-example} illustrates the semantics of \code{spread}.
\item The \code{gather} construct is the inverse of \code{spread}: It unpivots the input table by moving column names into the table body. Specifically, \code{gather} first eliminates all columns in $\bar{c}_\mathit{target}$, and then creates two new columns $c_\mathit{key}$ and $c_\mathit{val}$ where $c_\mathit{key}$ is filled with column names in $\bar{c}_\mathit{target}$ and column $c_\mathit{val}$ is filled with values in the eliminated columns. This is illustrated in the third drawing in \autoref{fig:ptbl-example}.
\item The \code{summarize} construct first partitions its input table into groups based on  values in $\bar{c}_\mathit{key}$  and then applies the function $\alpha$ to each group to aggregate values in column $c_\mathit{target}$. For example, in the rightmost part of \autoref{fig:ptbl-example}, the table is partitioned into two groups based on values in $c_1$ (labeled with different colors), and column $c_2$ is populated by taking the maximum of all values in the corresponding partition.
\end{enumerate}

\subsection{Visualization Language}
Our visualization language $\lang_{\viz}$ is shown in \autoref{fig:vis-lang}, which formalizes core constructs in Vega-Lite~\cite{DBLP:journals/tvcg/SatyanarayanMWH17}, the \code{ggplot2} visualization library for R and VizQL~\cite{DBLP:conf/sigmod/Hanrahan06} from Tableau. This formalization enables concise descriptions of visualizations by encoding data as properties of graphical marks. It represents single plots using a set of mappings that map data fields to visual properties, and supports combining single charts into compositional charts through layering and subplotting. A program $\pvis$ in this language  takes as input a table $\tab$ and outputs a visual trace $\trace$. Throughout this paper, we refer to programs in this language as \emph{visual programs}.

\begin{figure}[!t]
\[
\begin{array}{rlll}
\pvis & = &  \mathsf{MultiPlot}(SP, c_\mathsf{sub}) ~|~ SP \\
\mathit{SP} & = & \mathsf{MultiLayer}(\bar{L}) ~|~ L\\
\mathit{L} & = & \mathsf{Scatter}(c_x, c_y, c_\mathit{color}, c_\mathit{size}) & \text{(Scatter Plot)}\\
  & | & \mathsf{Line}(c_x, c_y, c_\mathit{color}) & \text{(Line Chart)}\\
  & | & \mathsf{Bar}(c_x, c_y, c_{y_2}, c_\mathit{color}) & \text{(Bar Chart)}\\
  & | & \mathsf{Stacked}(c_x, c_h, c_\mathit{color}) &  \text{(Stacked Bar Chart)} \\
c & = & \mathit{column} ~|~ \epsilon
\end{array}
\]
\vspace{-10pt}
\caption{The visualization language $\lang_{\viz}$.}
\label{fig:vis-lang}
\end{figure}

As shown in \autoref{fig:vis-lang}, a visual program $\pvis$ either creates a grid of multiple plots using the \textsf{MultiPlot} construct or  a single plot $SP$. Each  plot can in turn consist of multiple layers (indicated by the \textsf{MultiLayer} construct) or a single layer. Each layer is either a scatter plot (\textsf{Scatter}), a line chart (\textsf{Line}), a bar chart (\textsf{Bar}), or a stacked bar chart (\textsf{Stacked}).  The \textsf{MultiLayer} construct in this language is used to compose \emph{different} kinds of charts in the same plot (e.g., a scatter plot and a line chart), but our visualization language is nonetheless rich enough to allow layering the same type of chart within a plot: For example, the \textsf{Line} primitive can be used to render multiple line charts where each individual  line chart has a different color.

In terms of its semantics, a visual program $\pvis$ specifies how each tuple in the input table corresponds to a visual element in the output trace; thus, all constructs in  $\lang_{\viz}$  refer to column names in the input table. For instance, for the \textsf{MultiPlot} construct, the column name $c_\mathsf{sub}$ specifies that tuples sharing the same value of $c_\mathsf{sub}$ are to be visualized in the same subplot, whereas tuples with different values of $c_\mathsf{sub}$ belong to two different subplots. Similarly, for the \textsf{Line} construct, tuples that agree on the value of $c_\mathsf{color}$ are rendered as part of the same line chart, whereas tuples that disagree on the $c_\mathsf{color}$ value correspond to different layers.

\begin{figure}[!t]
\includegraphics[width=\textwidth]{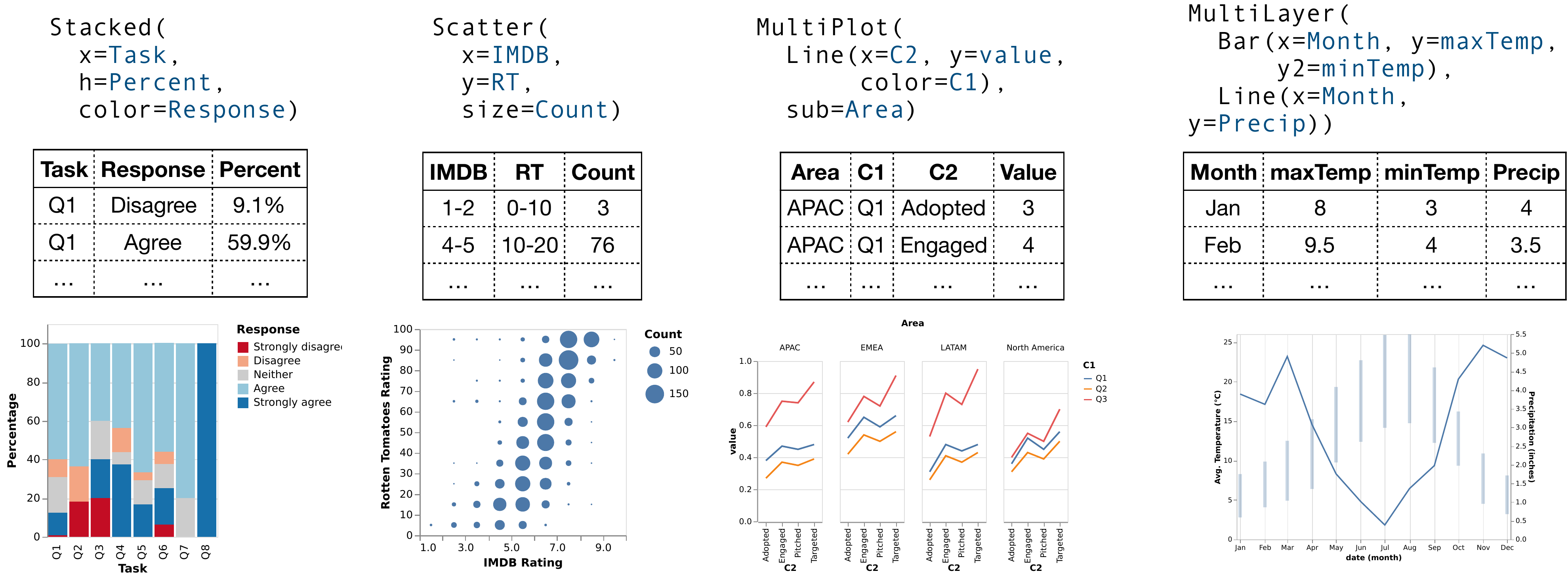}
\vspace{-10pt}
\caption{Examples of visualization operators in $\lang_\viz$ and their corresponding visualizations.}
\label{fig:pvis-example}
\end{figure}

In what follows, we explain the semantics of our visualization language with the aid of the examples shown in \autoref{fig:pvis-example}.

\begin{example}
The first example in \autoref{fig:pvis-example} shows a visual program for rendering  a stacked bar chart visualization of a study report. This  program specifies using \code{x=Task} that each different (stacked) bar on the $x$-axis should correspond to a different task (Q1, Q2 etc.) from the input table. The second argument, \code{y=Percent}, specifies that the height of each bar (within a stack) is determined by the \code{Percent} column in the input table. Finally, the third argument, \code{color=Response}, specifies that the color of each bar (within the stack) is determined by the value stored in the \code{Response} column of the input table.
\end{example}

\begin{example}
The second visual program in  \autoref{fig:pvis-example}  renders a scatter plot that visualizes the correlation between IMDB and Rotten Tomato reviews. Specifically, the \code{Scatter} construct draws a point for each row in the input table. In our example, the first  (resp. second) argument specifies that the $x$  (resp. $y$) coordinate is determined by the value in the \code{IMDB} (resp. \code{RT}) column of the input table. Finally, the third argument \code{size=Count} specifies that the size of the point is determined by the corresponding value stored in the \code{Count} column.
\end{example}

\begin{example}
The third program from \autoref{fig:pvis-example} renders multiple subplots, as specified by the \code{MultiPlot} construct. The second argument \code{subplot=Area}   specifies that each subplot corresponds to a separate value  in the \code{Area} column of the input table (APAC, North America etc.). The first argument, on the other hand, specifies that each subplot is a line chart. Furthermore, since the third argument of the \code{Line} construct is \code{color=C1}, each subplot consists of multiple line charts of different colors, determined by the value of the \code{C1} column in the input table.  Finally, the $(x,y)$ coordinates of the points within each line chart are determined by the values in \code{C2} and \code{Value} columns respectively.
\end{example}

\begin{example}
 The last program in \autoref{fig:pvis-example} draws a layered chart consisting of a bar chart and a line chart using the \textsf{MultiLayer} construct. Here,  bars show the temperature range for each month because the \code{x} value corresponds to the \code{Month} field in the input table, and \code{y} and \code{y2} correspond to the \code{maxTemp} and \code{minTemp} values for that month. On the other hand, the line chart shows the precipitation for each month; this is again specified using \code{x=Month} and \code{y=Precip}.
\end{example}

\section{Synthesis Algorithm}\label{sec:alg}
In this section, we first give an overview of our
top-level synthesis algorithm (\autoref{sec:alg-overview}) and then present techniques for learning visual programs (\autoref{sec:alg-viz}) and table transformation programs (\autoref{sec:alg-tbl}) respectively.

\begin{algorithm}[t]
  \caption{Top-level Synthesis Algorithm}\label{alg:synthesis}
  {
  \begin{algorithmic}[1]
  \Procedure{Synthesize}{$\tab_\text{in}$, $\vsketch$}
  \State {\rm \bf input:} Input table  $\tab_\text{in}$, visual sketch $\vsketch$
  \State {\rm \bf output:} A table transformation program $\ptbl$ and 
  visual program $\pvis$, or $\bot$ if failure
  \vspace{0.1in}
  \State $\Omega$ $\leftarrow$ \Call{LearnVisualProgs}{$\vsketch$}
  \ForAll{$(\pvis, \tblconstraint) \in \Omega$}  
      \State $r$ $\leftarrow$ \Call{LearnTableTransform}{$\intbl, \tblconstraint$}
      \State {\bf match} $r$
      \State \quad {\bf case} $\bot$: {\bf \ continue} 
      \State \quad {\bf case} $(\ptbl, \sigma)$: \Return $(\ptbl, \pvis[\sigma])$
  \EndFor
  \State \Return $\bot$
  \EndProcedure
  \end{algorithmic}
  }
\end{algorithm}

\subsection{Overview}~\label{sec:alg-overview}

Algorithm~\ref{alg:synthesis} describes our top-level visualization-by-example algorithm, which takes as input a table $\intbl$ and a visual sketch $\vsketch$ (expressed as a visual trace in the notation of \autoref{fig:trace-lang}) and returns a pair of programs $(\ptbl, \pvis)$ such that $\vsketch \subseteq \pvis(\ptbl(\intbl))$ (or $\bot$ to indicate failure). As mentioned previously, our synthesis algorithm is compositional in the sense that it uses an intermediate specification to guide the search for table transformation programs.

\algnewcommand\algorithmicforeach{\textbf{for all}}
\algdef{S}[FOR]{ForAll}[1]{\algorithmicforall\ #1\ \algorithmicdo}

Internally, the \textsc{Synthesize} procedure first uses the input visual sketch $\vsketch$ to infer a \emph{set} $\Omega$ of intermediate synthesis results. Each intermediate  result $r \in \Omega$ is a pair $(\pvis, \tblconstraint)$, where $\pvis$ is a visual program that is consistent with the provided visual sketch and $\tblconstraint$ is a constraint that imposes certain requirements on the input to $\pvis$. In other words, for a given visual program $\pvis$, $\tblconstraint$ serves as an intermediate specification that constrains the space of possible table transformation programs. However, since we do not know the column names used in this intermediate table, both $\pvis$ and $\tblconstraint$  refer to "made-up" column names to be resolved in the next phase.

In the second phase of the algorithm (lines 5-9), we try to find a table transformation program that satisfies the intermediate specification. Specifically, for each intermediate  synthesis result $(\pvis, \tblconstraint)$,  the {\sc LearnTableTransform} procedure is used to synthesize a table transformation program $\ptbl$ and a column mapping $\sigma$ such that $\ptbl(\intbl)$ satisfies the constraint $\tblconstraint[\sigma]$. Thus, if {\sc LearnTableTransform} does not return $\bot$ to indicate failure, the program $\pvis[\sigma] \circ \ptbl$ is guaranteed to satisfy the end-to-end specification defined by $(\intbl, \vsketch)$. 

\begin{figure}[!t]
\small

\AXC{$\trace_i \in \mathsf{partitionBySubplots}(\trace)$ \quad $\mathsf{c_{sub}} \ {\rm fresh}$ \quad $\trace_i \Uparrow (p, \psi_i)$ \quad $i \in [1,n]$}
\RightLabel{\ (Multi-Plot)}
\UIC{$\trace \Uparrow \left(\mathsf{MultiPlot}(p, \mathsf{c_{sub}}), \bigwedge_i\psi_i\right)$ }
\DP

\bigskip

\AXC{$\trace_i \in \mathsf{partitionByType}(\trace)$ \quad 
 $\trace_i \Uparrow (l_i, \psi_i)$ \quad  $i \in [1,n]$}
\RightLabel{\ (Multi-Layer)}
\UIC{$\trace \Uparrow \left(\mathsf{MultiLayer}(\bar{l}_i), \bigwedge_{i=1}^n\psi_i\right)$ } 
\DP

\bigskip

\AXC{
$\tab=\bigcup_{i=1}^n (a_{x}^i, a_{y}^i, a_\mathit{c}^i, a_\mathit{s}^i, a_\mathit{sub}) $ 
\quad \quad $\mathsf{c_x},\mathsf{c_y},\mathsf{c_{color}, \mathsf{c_{size}}}$ \ {\rm fresh}
}
\RightLabel{\ (Scatter)}
\UIC{
$\bigcup_{i=1}^n \left\{\mathsf{point}(a_{x}^i,a_{y}^i, a_\mathit{c}^i, a_\mathit{s}^i, a_\mathit{sub})\right\} 
\Uparrow \left(\mathsf{Scatter}(\mathsf{c_x},\mathsf{c_y},\mathsf{c_{color}}, \mathsf{c_{size}})\right), \tbl \substcontain t$ 
}
\DP

\bigskip

\AXC{
$\tab=\bigcup_{i=1}^n (a_{x}^i,a_{y}^i, a_{y_2}^i, a_\mathit{c}^i, a_\mathit{sub}) $ 
\quad \quad $\mathsf{c_x},\mathsf{c_y},\mathsf{c_{y_2}},\mathsf{c_{color}}$ \ {\rm fresh}
}
\RightLabel{\ (Simple Bar)}
\UIC{
$\bigcup_{i=1}^n \left\{\mathsf{bar}(a_{x}^i,a_{y}^i, a_{y_2}^i, a_\mathit{c}^i, a_\mathit{sub})\right\} 
\Uparrow \left(\mathsf{Bar}(\mathsf{c_x},\mathsf{c_y},\mathsf{c_{y_2}}, \mathsf{c_{color}})\right), \tbl \substcontain t$ 
}
\DP

\bigskip
\medskip

\AXC{
\stackunder
  {$\tab=\bigcup_{i=1}^n (a_{x}^i,a_{y_2}^i - a_{y}^i, a_\mathit{c}^i, a_\mathit{sub})$ 
        \quad \quad $\mathsf{c_x},\mathsf{c_h},\mathsf{c_{color}}$ \ {\rm fresh}}
  {$\psi_0=\forall i \in [1,n]. \sum_{r\in\{r\in\tab_\text{in}\mid r[\mathsf{c_x}]=a_x^i\land r[\mathsf{c_{sub}}]=a_\mathit{sub}\land r[\mathsf{c_{color}}]<a_\mathit{c}^i\}} r[\mathsf{c_h}]=a_{y}^i$}
}
\RightLabel{\ (Stacked Bar)}
\UIC{$\bigcup_{i=1}^n \left\{\mathsf{bar}(a_{x}^i,a_{y}^i, a_{y_2}^i, a_\mathit{c}^i, a_\mathit{sub})\right\}  \Uparrow \left(\mathsf{Stacked}(\mathsf{c_x},\mathsf{c_h}, \mathsf{c_{color}}), \psi_0\land \tbl \substcontain t \right)$ }
\DP

\bigskip
\medskip

\AXC{
\stackunder
  {$\tbl_1= \bigcup_{i=1}^n (a_{x_1}^i,a_{y_1}^i, a_{c}^i, a_{p})$ \quad $\tbl_2= \bigcup_{i=1}^n (a_{x_2}^i,a_{y_2}^i, a_{c}^i, a_{p})$ \quad $\mathsf{c_x},\mathsf{c_y},\mathsf{c_{color}}$ \ {\rm fresh}}
    {$\psi_0=\forall i \in [1,n].\not\exists r\in \intbl.~(r[\mathsf{c_{color}}]=a_{\mathit{c}}^i\land r[\mathsf{c_{\mathsf{sub}}}]=a_{p})\rightarrow a_{x_1}^i \le r[\mathsf{c_x}]\le a_{x_2}^i$}
}\RightLabel{\ (Line)}
\UIC{$\bigcup_{i=1}^n \left\{\mathsf{line}(a_{x_1}^i,a_{y_1}^i,a_{x_2}^i,a_{y_2}^i, a_{c}^i, a_{p})\right\} \Uparrow \left (\mathsf{Line}(\mathsf{c_x},\mathsf{c_y},\mathsf{c_{color}}), \tblconstraint_0 \land \tbl_1 \substcontain \intbl \land \tbl_2 \substcontain t \right)$ }
\DP
\caption{Inference rules describing synthesis of visual programs}
\label{fig:viz-bw-semantics}
\end{figure}

\subsection{Synthesis of Visual Programs}
\label{sec:alg-viz}
In this section, we describe the {\sc LearnVisualProgs} procedure used in Algorithm~\ref{alg:synthesis}. This procedure is described using inference rules of the form
$
\trace \Uparrow (\pvis, \tblconstraint)
$ where $\trace$ is a visual trace, $\pvis$ is a visual program, and $\tblconstraint$ is a constraint. 
The meaning of this judgment is that, if the input table $\tbl$ satisfies  constraint $\tblconstraint$, then $\pvis(\tbl)$ yields a visualization that is consistent with $\trace$ (i.e., $\trace \subseteq \pvis(\tbl)$). Observe that, for a given visual trace $\trace$, there may be multiple programs that are consistent with it --- i.e., we can have $\trace \Uparrow (\pvis^i, \psi^i)$ for multiple values of $i$. This is the reason why the {\sc LearnVisualProgs} procedure used in \autoref{alg:synthesis} returns a set rather than a singleton. In what follows, we explain each of the inference rules from \autoref{fig:viz-bw-semantics} in more detail.

\paragraph{Multiple plots.} The first rule, labeled Multi-Plot, is used to synthesize programs for generating multiple subplots. Since each element in a visual trace has an attribute that identifies which subplot it belongs to, we first partition the visual elements according to the value of this attribute. This allows us to obtain $n$ different visual traces $\trace_1, \ldots, \trace_n$, and we recursively synthesize a visual program $p_i$ and a constraint $\tblconstraint_i$ for each visual trace $\trace_i$. However, since the \textsf{MultiPlot} construct takes a \emph{single} program as argument, this means that all  subplots must be generated using the \emph{same} visual program; thus, the premise of this rule stipulates that all $p_i$'s must be the same program $p$. On the other hand, each subplot can  impose different restrictions on the input table; thus, the input  has to satisfy all of these constraints (i.e.,  $\bigwedge_{i=1}^n \tblconstraint_i$). 
Finally, since we do not know which column of the input table is used to generate different subplots, we make up a fresh column name called $\mathsf{c_{sub}}$ and return the synthesized program \textsf{MultiPlot}($p, \mathsf{c_{sub}}$) as the solution.

\paragraph{Multiple layers.} The second rule, Multi-Layer, is similar to the Multi-Plot rule and is used to generate programs that compose different types of charts. Similar to the previous rule, we again partition elements in the visual trace according to their type (i.e., point, bar etc.) to obtain $n$ different traces $\trace_1, \ldots, \trace_n$ and recursively synthesize a visual program $l_i$ and a constraint $\tblconstraint_i$ for each $\trace_i$. Then, the synthesized program $\mathsf{MultiLayer}(\myvec{l})$ will generate a visualization consistent with the visual sketch as long as the input table satisfies $\bigwedge_{i=1}^n \tblconstraint_i$.

\paragraph{Scatter plot.} The next rule is used to synthesize a visual program that renders  a scatter plot. Since all elements in a scatter plot must be points, the precondition of this rule requires that the visual trace is a set of points with the same subplot attribute. Furthermore, for each point $p$ with attributes $\myvec{a}^i$ in the visual sketch, there must be a corresponding row in the input table that contains exactly the values $\myvec{a}^i$. To express this requirement on the input table, we construct a table  $\tbl$ that contains rows $\myvec{a}^i$ and generate the constraint $\tbl \substcontain t$ where $t$  refers to the input table for the synthesized visual program. Finally, since we do not know the names of the columns in the input table, we introduce placeholder column names $\myvec{c}$ and return  the program $\mathsf{Scatter}(\myvec{c})$.

\paragraph{Bar charts.} The next two rules, labeled Simple Bar and Stacked Bar, both generate bar charts and are very similar to the previous Scatter rule. For Stacked Bar,  $\mathsf{c_h}$ represents  the height of the bar rather than the absolute $y$-position; thus,
we compute entries in column $\mathsf{c_h}$ as $a^i_{y_2}-a^i_y$ for the $i$'th row in the table sketch. Also, in addition to the constraint $\tbl \substcontain t$, the Stacked Bar rule imposes an additional constraint on the input table. In particular, since the bars in a Stacked Bar chart must be stacked directly on top of each other, constraint $\psi_0$ essentially stipulates that the starting $y$ position of one bar is  precisely the end $y$-position of the previous stack below it. In practice, when computing constraint $\psi_0$, we compute the end $y$ position of the  stack below by summing the heights of all the individual bars below the current one.

\paragraph{Line chart.} The final rule is used to synthesize a \textsf{Line} program in our visualization language. Recall that a line visual element is defined by its two end points $(a_{x_1}, a_{y_1})$ and $(a_{x_2}, a_{y_2})$, and these end points must correspond to two different rows in the input table. Thus, we generate two different constraints  $\tbl_1 \substcontain t$ and $\tbl_2 \substcontain t$ that describe requirements imposed by the left end and  right end of each line segment respectively. Finally, the constraint $\psi_0$ in the second line of the premise imposes the following additional restriction: If the visual sketch contains a line segment with $(a_x, a_y)$ and $(a_x',a_y')$ as its end points, there should not be another entry in the input table that belongs to the same line chart (i.e., same color and subplot) but where the $x$ value is in the range $(a_x, a_x')$. Without this additional constraint $\psi_0$, the generated visualization would {not} be guaranteed to satisfy the  provided  visual sketch.

\paragraph{Properties} Our visual program inference procedure enjoys the following properties that are important for the soundness and completeness for the overall approach.

\begin{property}[Decomposition] Suppose that $\trace\Uparrow (\pvis,\tblconstraint)$ and  $\tbl$ is a table that satisfies constraint $\tblconstraint$ (i.e., $\tbl \models \tblconstraint$). Then, we have $\trace \subseteq \pvis(\tbl)$.
\end{property}

The above property shows the soundness of the overall the synthesis algorithm. In particular, let $\pvis$ be a visual program synthesized in the first phase. Based on the above property, as long as we can find a table transformation program $\ptbl$ that satisfies the specification $(\intbl, \tblconstraint)$, then we are guaranteed that the composition $\pvis \circ \ptbl$ will satisfy the specification of the overall synthesis task.

\begin{property}[Completeness]  Let $\trace$ be a visual sketch, and suppose that there exists a table $\tbl$ and a visual program $\pvis$ in $\lang_\viz$ such that $\trace \subseteq \pvis(\tbl)$. Then, we have $\trace \Uparrow (\pvis, \tblconstraint)$ such that $\tbl \models \tblconstraint$.
\end{property}

This second property shows the completeness of the overall synthesis algorithm. In particular, it states that, if there exists a table $\tbl$ and visual program $\pvis$ such that $\pvis(\tbl)$ is consistent with the given visual sketch, then our inference procedure will (a) return $\pvis$ as one of the solutions, and (b) $\tbl$ will satisfy the  constraint $\tblconstraint$ associated with $\pvis$.

\subsection{Synthesizing Table Transformations via Bidirectional Reasoning}~\label{sec:alg-tbl}

In this section, we describe the {\sc LearnTableTransform} function used in \autoref{alg:synthesis}. This procedure is given in \autoref{alg:table} and takes as input the original input table $\intbl$ and the intermediate specification $\outconstraint$  generated during the first phase.   {\sc LearnTableTransform} either returns a program $\ptbl$ such that $\ptbl(\intbl)$ is consistent with the specification $\outconstraint$ or yields  $\bot$ to indicate failure. If synthesis is successful, {\sc LearnTableTransform} additionally returns a mapping $\sigma$ from the made-up column names used in $\outconstraint$ to the actual column names used in  $\ptbl(\intbl)$.

\begin{algorithm}[ht]
  \begin{algorithmic}[1]
\Procedure{LearnTableTransform}{$\intbl, \outconstraint$}
\State $\inconstraint \gets (t_0 \subseteq \intbl \land \intbl \subseteq t_0)$
\While{\textsf{existsNextSketch}()}
\State $\partialprog_0 \gets \mathsf{getNextSketch}();$ 
  \State $\worklist \gets \{ (\partialprog_0, \sigma) \ | \ \sigma \in \mathsf{Mappings}(\mathsf{Cols}(\outconstraint), \mathsf{Cols}(\partialprog_0) \};$

  \While{$\neg \worklist.\mathsf{isEmpty}()$}
    \State $(\partialprog, \sigma) \gets \worklist.\mathsf{next}()$
    \If{$\mathsf{IsComplete}(\partialprog)$ }
    \If{$\partialprog(\intbl) \models \outconstraint[\sigma]$}  \Return $(\partialprog, \sigma)$ 
    \Else \ {\bf continue;}\EndIf
    \EndIf
    \State $\phi \gets \mathsf{Analyze}^+(\inconstraint, \partialprog)\land \mathsf{Analyze}^-(\outconstraint[\sigma], \partialprog);$
    \If{$\mathsf{UNSAT}(\phi)$} $\mathbf{continue}$ \EndIf
    \State  $\square_k \gets \mathsf{chooseHole}(\partialprog)$ 
  \State $\worklist \gets \worklist \cup \left\{ (\partialprog[\square_k \mapsto v], \sigma') \ \big{|} \ v \in \mathsf{dom}(\square_k), \mathsf{Mappings}(\mathsf{Cols}(\outconstraint), \mathsf{Cols}(\partialprog) \cup \{ v \} \right\}$
 
  \EndWhile
\EndWhile
\Return $\bot$
\EndProcedure
  \end{algorithmic}
  \caption{Table transformation synthesis algorithm. }
  \label{alg:table}
\end{algorithm}

From a high level, the outer loop of \autoref{alg:table} lazily enumerates program sketches based on the language from \autoref{sec:table-dsl}. In this context, a program sketch  $\partialprog$ is a sequence of instructions of the form $t = \code{op}(\square_1, \ldots, \square_n)$ where $t$ is a program variable, \code{op} is a construct  in the table transformation language (e.g., \code{mutate}, \code{join}), and each $\square_i$ is a \emph{hole} representing an unknown argument. To obtain a program that is a completion of $\partialprog$, we need to fill each of the holes in the sketch with  previously defined program variables or column names from the input table.

In more detail, the algorithm maintains a worklist $\worklist$ of elements $(\partialprog, \sigma)$ where  $\partialprog$ is a (partially completed) program sketch and $\sigma$ is a possible mapping from the made-up column names in $\outconstraint$ to actual column names in the output table.  In particular, $\sigma$  maps each column name used in $\outconstraint$ to an element in $\mathsf{Cols}(\partialprog)$, where $\mathsf{Cols}(\partialprog)$ includes both the columns used in $\intbl$  as well as any additional columns mentioned in $\partialprog$. In each iteration of the inner while loop, the algorithm dequeues (at line 8) a pair $(\partialprog, \sigma)$ and checks whether $\partialprog$ is a complete program (i.e., no holes). If this is the case and $\partialprog$ satisfies the specification under mapping $\sigma$ (line 9), we then return $(\partialprog, \sigma)$  as a solution. On the other hand, if $\partialprog$ contains any remaining holes, we  perform bidirectional program analysis (line 11) to check if there is any completion of $\partialprog$ that \emph{can} satisfy the (intermediate) specification $\outconstraint$. In particular, line 11 of \autoref{alg:table} generates a constraint $\phi$ that is a conjunction of atomic predicates of the form $e_1 \subseteq^* e_2$ where $ \subseteq^*$ represents  the table inclusion relations defined earlier  ($\subseteq$ , $\substcontain$), and each $e_i$ is either a program variable or a concrete table. If these generated constraints result in a contradiction, there is \emph{no} sketch completion that is consistent with $\outconstraint$; thus, the algorithm moves on to the next element in the worklist (line 12). On the other hand, if we cannot prove the infeasibility of $\partialprog$, we pick one of the holes $\square_k$ used in the sketch and add a new set of partial programs to the worklist by instantiating that hole with some element in its domain (line 13). The domain of the hole is determined by its type, columns in the input schema for the given statement, and previously defined variables in the partial program. Since hole $\square_k$ may have been filled with a new column name $v \not \in {\rm Cols}(\partialprog)$, we therefore also update the  worklist to consider any new mappings $\sigma'$ that we have not previously considered.

\begin{figure}[!t]
\small

\bigskip

\AXC{$\phi\Rightarrow t \subseteq^*\tab$}
\UIC{$\phi \forward t'=\mathsf{filter}(t, \_) : \phi\land (t' \subseteq^* \tab)$}
\DP
\qquad
\AXC{$\phi\Rightarrow t\subseteq^*\tab$}
\UIC{$\phi \forward t'=\mathsf{select}(t, \_) : \phi\land (t'\subseteq^*\tab)$}
\DP

\medskip
\bigskip

\AXC{$\phi\Rightarrow t\subseteq^*\tab$ \quad $\tab'=\eval{\mathsf{mutate}(\tab, c_t, \mathit{op}, \bar{c})}$}
\UIC{$\phi \forward t'=\mathsf{mutate}(t, c_t, \mathit{op}, \bar{c}): \phi\land (t'\subseteq^*\tab')$}
\DP
\qquad
\AXC{$\phi\Rightarrow (t_1\subseteq^*\tab_1 \land t_2\subseteq^*\tab_2)$}
\UIC{$\phi \forward t'=\mathsf{join}(t_1, t_2, \_) : \phi\land (t'\subseteq^* \tab_1\times\tab_2)$}
\DP

\medskip
\bigskip

\AXC{$\phi\Rightarrow t\subseteq^* \tab$ \quad $\tab'=\eval{\mathsf{gather}(\tab, \bar{c}_\mathit{id}, \bar{c}_\mathit{target})}$}
\UIC{$\phi \forward t'=\mathsf{gather}(t, \bar{c}_\mathit{id}, \bar{c}_\mathit{target}) : \phi\land (t'\subseteq^* \tab')$}
\DP
\qquad
\AXC{$\phi\Rightarrow t\subseteq^* \tab$ \quad $\tab'=\eval{\mathsf{spread}(\tab, \bar{c}_\mathit{id}, c_\mathit{key}, c_\mathit{val})}$}
\UIC{$\phi \forward t'=\mathsf{spread}(t, \bar{c}_\mathit{id}, c_\mathit{key}, c_\mathit{val}) : \phi\land (t'\substcontain \tab')$}
\DP

\medskip
\bigskip

\AXC{$\forall i\in[1,n].~\phi_{i-1} \forward t_i=e_i : \phi_i$}\RightLabel{(Chain)}
\UIC{$\phi_0 \forward \{t_1=e_1;\dots;t_n=e_n\}: \phi_n$}
\DP

\caption{Forward inference. Operator ``$\subseteq^*$'' refers to either $\subseteq$ or $\mathrel{\stackon[1pt]{$\subseteq$}{$\scriptstyle\diamond$}}$. In cases where the premise of no rule matches, we have an implicit judgment $\phi \forward s: \phi$ to propagate the input constraint. }
\label{fig:forward}
\end{figure}

As is evident from the above discussion, a key part of our table transformation synthesis algorithm is the $\mathsf{Analyze}^+$ and $\mathsf{Analyze}^-$ procedures for performing forward and backward inference to generate table inclusion constraints. These procedures are described in \autoref{fig:forward} and \autoref{fig:backward} using inference rules of the form $\phi \downarrow s: \phi'$ (for the forward analysis) and $\phi \uparrow s: \phi'$ (for the backward  analysis). The meaning of the judgment $\phi \downarrow s: \phi'$ is that, assuming $\phi$ holds \emph{before} executing statement $s$, then $\phi'$ must  hold \emph{after} executing $s$. Similarly, $\phi \uparrow s: \phi'$ means that, if $\phi$ holds \emph{after} executing $s$, then $\phi'$ must hold \emph{before} $s$ (i.e., $\phi'$ is a \emph{necessary} precondition for $\phi$ but may not be \emph{sufficient} to guarantee it). Since the inference rules shown in \autoref{fig:forward} and~\autoref{fig:backward} follow from the semantics of the table transformation language, we do not explain them in detail.  However, a key design decision  is that our analysis \emph{on purpose} does not compute  strongest post-conditions (for the forward analysis) or  strongest necessary preconditions (for the backward analysis) in order to ensure that the cost of deductive reasoning does not overshadow its benefits.  For example, in the reasoning rule for \code{summarize} in \autoref{fig:backward}, we over-approximate all aggregation functions as uninterpreted functions; thus the inferred pre-condition only requires that  input table $t$  include content from non-aggregated columns ($\tab'$) in the output table $t'$. While a more precise analysis rule could consider the underlying semantics of different aggregation operators, this kind of reasoning would be prohibitively expensive~\cite{DBLP:journals/pacmpl/WangCB18} and outweigh the benefits obtained from better pruning.

For the same reason, our procedure for checking satisfiability of table inclusion constraints (described in \autoref{fig:table-inclusion-dp}) is also  incomplete and intentionally over-approximates satisfiability. Thus, while the unsatisfiability of the generated constraints ensures the infeasibility of a given partially completed sketch, the converse is not true -- that is, our deductive reasoning technique may fail to prove infeasibility of a sketch even though no valid completion exists.

\begin{figure}
\small

\AXC{$\phi \Rightarrow \tab\substcontain t'$}
\UIC{$\phi \backward t'=\mathsf{filter}(t, \_): \phi\land (\tab\substcontain t)$}
\DP
\qquad
\AXC{  
$\phi\Rightarrow \tab\substcontain t'$ \quad $\tab'=\tab[-c_\mathit{target}]$}
\UIC{$\phi \backward t'=\mathsf{mutate}(t, c_\mathit{target}, \_, \_): \phi\land (\tab'\substcontain t)$}
\DP

\bigskip

\AXC{$\phi \Rightarrow \tab\substcontain t'$}
\UIC{$\phi \backward t'=\mathsf{select}(t, \_): \phi\land (\tab\substcontain t)$}
\DP
\qquad
\AXC{$\phi\Rightarrow \tab\substcontain t'$ \quad $\tab'=\eval{\mathsf{gather}(\tab, \bar{c}_\mathit{id}, \mathsf{schema}(\tab)-\{\bar{c}_\mathit{id}\})}$}
\UIC{$\phi \backward t'=\mathsf{spread}(t, \bar{c}_\mathit{id}, \_,\_) : \phi\land (\tab'\substcontain t)$}
\DP

\bigskip

\AXC{$\phi\Rightarrow \tab\substcontain t'$ \quad $\tab'=\mathit{RemoveDuplicates}(\tab[\bar{c}_\mathit{id}])$}
\UIC{$\phi \backward t'=\mathsf{gather}(t, \bar{c}_\mathit{id}, \_) : \phi\land (\tab'\substcontain t)$}
\DP
\quad
\AXC{$\phi\Rightarrow \tab\substcontain t'$ \quad $\tab'=\tab[-c_\mathit{target}]$}
\UIC{$\phi \backward t'=\mathsf{summarize}(T, \_, \_, c_\mathit{target}) : \phi\land (\tab'\substcontain t)$}
\DP

\bigskip

\AXC{$\forall i\in[1,n].~\phi_i\backward t_i=e_i : \phi_{i-1}$}\RightLabel{(Chain)}
\UIC{$\phi_n \backward \{t_1=e_1;\dots;t_n=e_n\}: \phi_0$}
\DP

\caption{Backward inference. Operator ``$\subseteq^*$'' refers to either $\subseteq$ or $\mathrel{\stackon[1pt]{$\subseteq$}{$\scriptstyle\diamond$}}$, and \emph{RemoveDuplicates}  removes duplicate tuples from the input table. As in the forward analysis, we assume an implicit rule $\phi \backward s: \phi$ that applies if none of the other premises are met.}
\label{fig:backward}
\end{figure}

\begin{figure}
\AXC{$\phi=e_1\subseteq^*e_2\land\phi_0$}
\UIC{$\phi\Rightarrow e_1\subseteq^*e_2$}
\DP
\quad
\AXC{$\phi\Rightarrow e_1\subseteq^*e_2$ \quad $\phi\Rightarrow e_2\subseteq^*e_3$}
\UIC{$\phi\Rightarrow e_1\subseteq^*e_3$}
\DP
\quad
\AXC{$\phi\Rightarrow e_1\subseteq e_2$}
\UIC{$\phi\Rightarrow  e_1 \substcontain e_2$}
\DP

\bigskip

\AXC{$\phi\Rightarrow T_1 \subseteq^* t$ \quad $\phi\Rightarrow t \subseteq^* T_2$ \quad $T_1 \not \subseteq^* T_2$ }
\UIC{$\phi\Rightarrow  \bot$}
\DP
\caption{Inference rules for checking satisfiability of table inclusion constraints. Operator ``$\subseteq^*$'' refers to either $\subseteq$ or $\mathrel{\stackon[1pt]{$\subseteq$}{$\scriptstyle\diamond$}}$, and metavariable $e$ refers to either a program variable $t$ in $\phi$ or a concrete table $\tab$.}
\label{fig:table-inclusion-dp}
\end{figure}

\paragraph{Properties} We end this section by describing some salient properties of \autoref{alg:table} that are important for the soundness and completeness of the end-to-end synthesis approach.

\begin{property}[Forward Analysis] Let $\partialprog$ be a partially completed sketch with argument $t$ and return parameter $t'$. Then, if  $t = \intbl \forward \partialprog:  \phi$ and $\phi \Rightarrow (t' \substcontain \tbl')$, then  \emph{any} completion $\prog$ of $\partialprog$ satisfies $\prog(\intbl) \substcontain \tbl'$.
\end{property}

By design, our forward analysis rules exploits the fact that many table transformation operators are monotonic over the input. This property essentially captures the correctness of forward inference. In particular, it says that, if we deduce that the output of $\partialprog$ on $\intbl$ is a sub-table of $\tbl'$, then this is true for every completion of $\partialprog$. The following property states something similar for the backward analysis:

\begin{property}[Backward Analysis] Let $\phi$ be a constraint and $\partialprog$ be a partially completed sketch with input parameter $t$.  Then, if $\phi \uparrow \partialprog : \phi'$ and $\phi'\Rightarrow (\tab'\substcontain t)$, then for any completion $\prog$ of $\partialprog$ and  any input table ${\tab}$ such that $\prog(\tab) \models \phi$, we have $\tab'\substcontain\tab$.
\end{property}

Similarly, our backward analysis rules by design conservatively propagate known values from the output to inputs. This property states that any conclusions reached by backward inference apply to all completions of $\partialprog$. Finally, we can state the following property about the correctness of our pruning strategy:

\begin{property}[Pruning Soundness] Given a partially completed sketch $\partialprog$, suppose we have  $\psi\backward \partialprog : \phi^-$ and $(t=\tab_\text{in})\forward \partialprog : \phi^+$. Let $P$ be a completion of $\partialprog$ such that $P(\intbl) \models \psi$, and let $\sigma$ be the resulting valuation after executing $P$ on $\intbl$. Then, we have $\sigma \models \phi^+ \land \phi^-$.
\end{property}

In other words, our pruning technique never rules out completions of $\partialprog$ that actually satisfy the given specification $(\intbl, \psi)$.

\section{Implementation}\label{sec:impl}

We have implemented the proposed technique in a tool called \toolname, which is written in Python. \toolname takes two inputs, namely the original data source (which can consist of one or more tables) as well as a visual sketch. Currently, \toolname requires the visual sketch to be expressed as a visual trace; however, with some additional engineering effort, it would be possible to integrate \toolname with  visual demonstration interfaces such as
Lyra~\cite{lyra} or VisExemplar~\cite{VisExemplar} to automatically generate visual traces from the demonstration.

\paragraph{Extension to visualization Language} \toolname can generate visual programs in  Vega-Lite~\cite{DBLP:journals/tvcg/SatyanarayanMWH17}, ggplot2, and a subset of \code{Matplotlib}. To handle all of these libraries, our implementation supports a richer visualization DSL than the one given in  \autoref{fig:vis-lang}.  In particular, \toolname supports two additional visualization constructs, \code{AreaChart} and \code{StackedAreaChart}, which provide another mechanism for visualizing quantities that change over time. To support this richer visualization language, we also extend our visual trace language from  \autoref{fig:trace-lang} with an element called \code{area}. Besides adding new constructors, we also extend existing constructors to take additional attributes as input. For example, the attribute $a_\mathit{shape}$ allows specifying the shape associated with each point in a scatter plot. 
Another attribute, $a_\mathit{order}$, for line charts allows  specifying a custom order instead of using the default $x$-axis value.

\paragraph{Extension to table transformation Language} The table transformation language used in our implementation extends \autoref{fig:tbl-lang} with a few additional constructs inspired by commonly used operators in the \code{tidyverse} $R$ package. For example, the table transnformation DSL in our implementation allows another construct called \code{separate} that is commonly used for table reshaping as well as a construct called \code{cumsum} for computing cumulative sum for a given column. Our implementation also allows a  more expressive version of the  \code{mutate} construct that supports a broader set of binary computations including arithmetic operations and string concatenation. 

\paragraph{Multi-layered visualizations} To simplify presentation in the technical section, we assumed that a visual program takes a single table as input. However, in many visualization libraries (e.g., \code{ggplot2}), the semantics of the \code{MultiLayer}($l_1, \ldots, l_n$) construct is that each different nested visual program $l_i$ operates on the $i$'th input table. To support these richer semantics, our implementation  synthesizes multiple different table transformation programs for each layer. To achieve this goal, the inference procedure for visual programs generates $n$ different intermediate specifications, one for each layer in the visual program, and we use the same table transformation procedure to synthesize $n$ different programs.

\paragraph{Ranking}
Following the Occam's razor principle, \toolname explores programs in increasing order of program size up to a fixed bound $K$. In practice, to leverage the inherent parallelism of our algorithm, \toolname uses multiple threads to search for solutions of different sizes and ranks programs according to their size.

\section{Evaluation}\label{sec:eval}

In this section, we evaluate the effectiveness of our approach on 83 real world visualization tasks collected from online forums and tutorials for advanced users.
The goal of our evaluation is to examine the following research questions:
\begin{enumerate}
	\item Can \toolname   solve real-world visualization tasks based on small visual sketches?
	\item Does the decomposition of the synthesis task into two sub-problem improve synthesis efficiency? 
	\item Are there any advantages to using our proposed table transformation algorithm  compared to re-using an existing state-of-the-art technique?
\end{enumerate}

\subsection{Benchmarks} 
\label{sec:benchmarks}
We evaluate \toolname on 83 visualization benchmarks~\footnote{Available at \url{https://chenglongwang.org/falx-project}}, 63 of which are collected from highly-reputed visualization tutorials for Excel~\footnote{\url{https://sites.google.com/site/e90e50/},~\url{https://chandoo.org/wp/category/visualization/},~\url{https://peltiertech.com/}} and Vega-Lite~\footnote{\url{https://vega.github.io/vega-lite/examples/}}, and 20 of which are collected from the \code{ggplot2} sub-forum on StackOverflow. 
To collect these benchmarks, we went through a few hundred visualization examples and retained all tasks that conform to the following criteria:

\begin{enumerate}
    \item The target visualization is expressible in our language. (We note that more than 80\% of these visualizations are expressible in our language and discuss the remaining ones in \autoref{sec:limitations}.)
    \item The example contains the actual input table.
    \item The task requires some form of table transformation  to generate the intended visualization.
    \item There is a way to produce the target visualization based on information in the example.
\end{enumerate}

We have these criteria because (1) tasks that cannot be achieved using our visualization language are out of scope for this work, (2) we need the raw data as an input to our tool, (3) we do not want to evaluate on trivial benchmarks, and (4) we need the target visualization to determine if our tool can produce the correct program. 

Among our $83$ benchmarks, $40$ of them contain subplots or multi-layered charts. Furthermore, for most benchmarks, the original data source consists of a single table whose size ranges from $4\times3$ to $3686\times9$, with average size $100\times10$.

\subsection{Key Results}~\label{sec:key-results}

To evaluate \toolname on these benchmarks, we programmatically generated  small visual sketches consisting of $4$ randomly sampled visual elements per layer. Specifically, given a  target visualization expressed in our visual trace language, we  sampled $4$ elements from the corresponding visual trace and used this as our visual sketch. While the number $4$ is somewhat arbitrary, we believe that a visual sketch with four elements is small enough  that it would not be too onerous for users to construct such a sketch.  

Given these randomly generated  visual sketches, we evaluated \toolname using the following methodology. We fixed a time budget $t$, and we let \toolname explore multiple visualization scripts consistent with the  provided sketch within this time budget. Then, for a given value of $t$, we consider the benchmark to be solved  if any of the programs explored by \toolname within that time budget generates the intended visualization. 

\begin{table}[ht]
    \begin{minipage}{.45\linewidth}
      \centering
      \caption{Summary of experimental results.}
      \vspace{-5pt}
      \begin{tabular}{|c || c |}
        \hline
        Time budget & \# solved \\\hline
        1s & 26   \\
        10s & 49 \\
        60s & 62 \\
        600s & 70 \\\hline
        \end{tabular}
        \label{tbl:time}
    \end{minipage}%
    \begin{minipage}{.55\linewidth}
      \centering
      \caption{Impact of size of visual sketch on the ranking.}
      \vspace{-5pt}
      \begin{tabular}{|c || c | c | c | c|}
        \hline
        \# samples & top-1 & top-5 & top-10 & >10 \\\hline
        1 & 14 & 29 & 36 & 66 \\
        2 & 20 & 37 & 43 & 68 \\
        3 & 22 & 45 & 53 & 69 \\
        4 & 26  & 49 & 57 & 70 \\
        6 & 31 & 52& 57 & 70\\
        8 & 30 & 58 & 63 & 70\\\hline
        \end{tabular}
        \label{tbl:size-impact}
    \end{minipage} 
\end{table}

The results of this experiment are summarized in \autoref{tbl:time}. For a time budget of $600$ seconds, \toolname is able to solve $70$ out of $83$ benchmarks ($84\%$). If we reduce the time budget to a minute, then \toolname can solve $75\%$ of the benchmarks. Furthermore, $59\%$  of the benchmarks can be solved within 10 seconds, and $31\%$ can be solved within one second.

\autoref{tbl:size-impact} explores the same experimental data from a different perspective. Specifically, given a value $k$, let us consider a benchmark to be "solved" if the desired visualization is one of the first-$k$ visualizations returned by the tool. As shown in the first row of \autoref{tbl:size-impact}, among the 70 benchmarks that can be solved within the $600$ second time limit, 26 ($37\%$) of them  are ranked as the top-1 solution, and $49$ ($70\%$) and $57$ ($81\%$) are ranked as top-5 and top-10 respectively. Given that a user can quickly look through 10 visualization results and decide if any of them is the desired visualization, we believe these results affirmatively answer our first research question.

In \autoref{tbl:size-impact}, we also explore the impact of  sketch size on synthesis results. Specifically, recall that we generate the visual sketches by randomly sampling $n$ elements from the target visualization, and, so far, our discussion focused on the results for $n = 4$. \autoref{tbl:size-impact} shows the ranking of the desired visualization as we increase $n$ to $6$ and $8$ and decrease to $1$, $2$, $3$ respectively. For cases with more visual trace samples, since the visual sketch contains more information as we increase $n$, \toolname synthesizes fewer spurious programs and the ranking of the target visualization improves as a result.~\footnote{The reader may notice that $n=6$ does better compared to $n=8$ for the top-$1$ result; this is caused by the random sampling of visual elements.} This finding indicates that users can incrementally add more visual elements to the output example and gradually refine the synthesis results when the initial top-ranked solutions fail to meet the user's expectation.

\subsection{Evaluating Impact of Decomposition}

As mentioned throughout the paper, a key design choice underlying our technique is to decompose the visualization task by inferring an intermediate specification for each possible visual program. In this section, we aim evaluate the empirical significance of this decomposition. 

\begin{figure*}[t]
    \centering
    \begin{subfigure}{.5\textwidth}
      \centering
      \includegraphics[width=.8\linewidth]{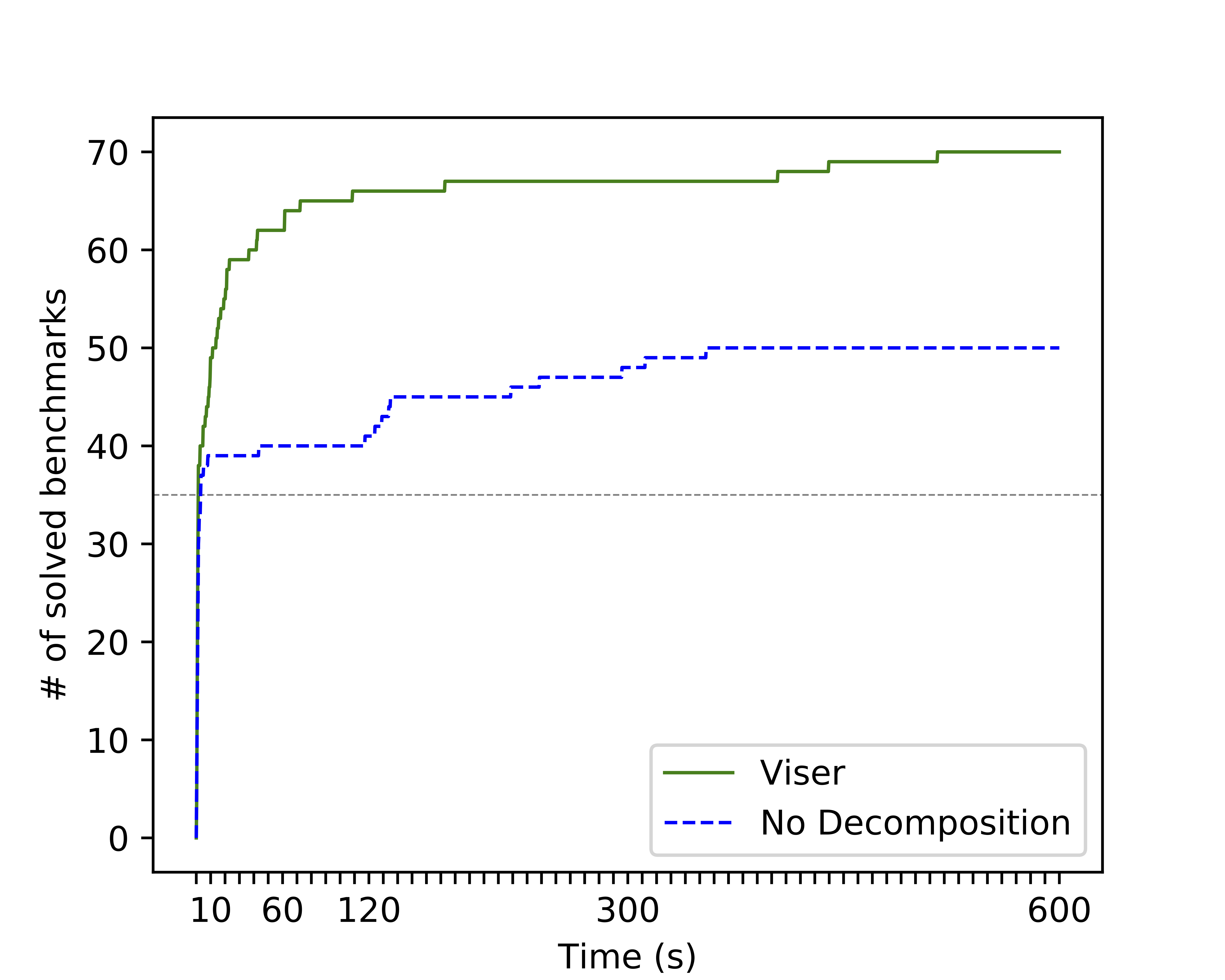}
      \caption{Comparison against baseline with no decomposition}
      \label{fig:eval-nodecompose}
    \end{subfigure}%
    \begin{subfigure}{.5\textwidth}
      \centering
      \includegraphics[width=.8\linewidth]{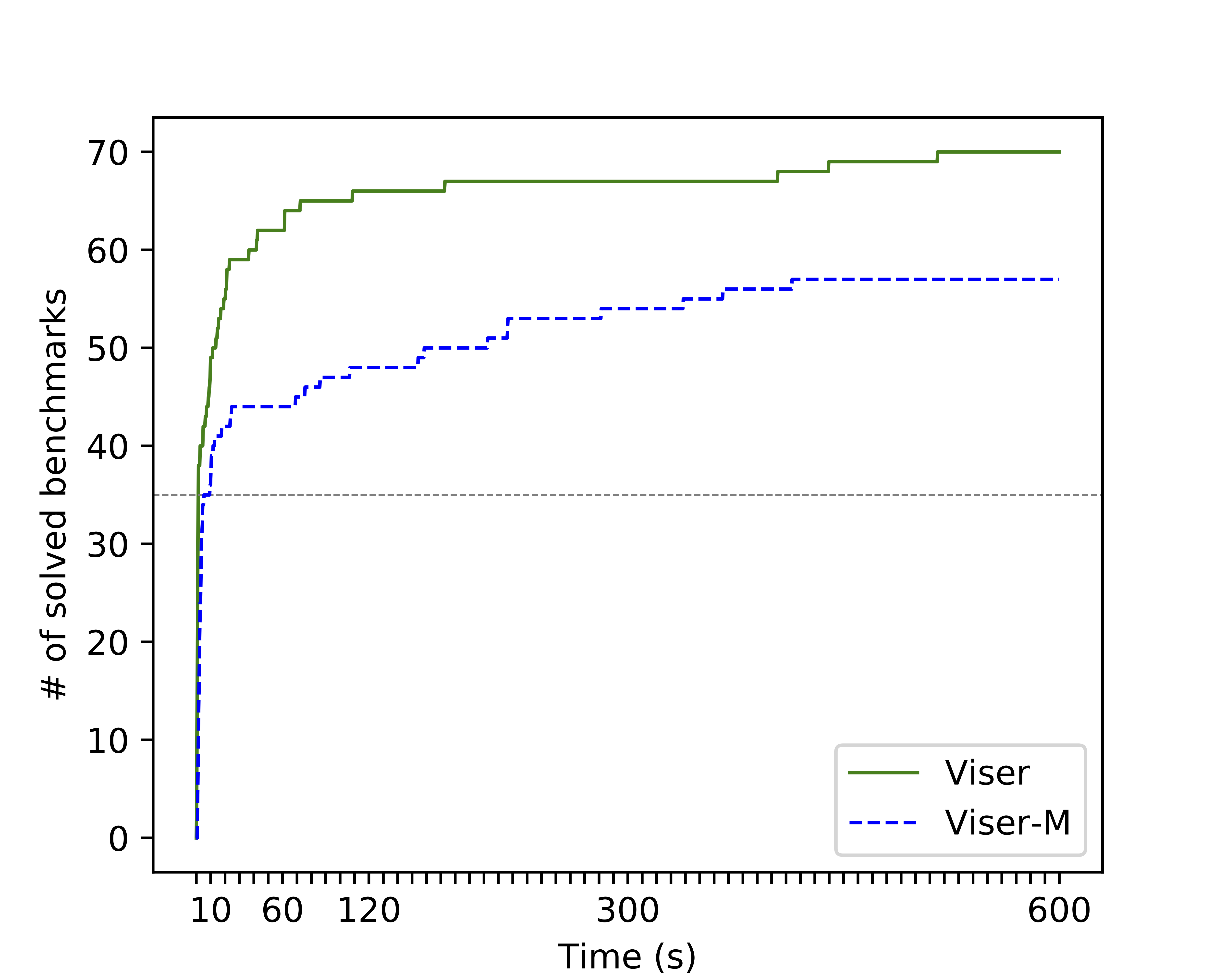}
      \caption{Comparison against {\sc \toolname-M}}
      \label{fig:eval-morpheus}
    \end{subfigure}
    \caption{Comparison of \toolname against different baselines}
\end{figure*}

To perform this study, we implement a baseline using the following methodology: Similar to \toolname, the baseline first infers a visual program $\pvis$ consistent with the sketch as discussed in \autoref{sec:alg-viz}; however, the baseline approach does not generate an intermediate specification. Then, during table transformation synthesis, for every enumerated program $\ptbl$, the baseline checks whether $\pvis(\ptbl(\intbl))$  is consistent with the visual sketch. In other words, without the intermediate specification, table transformation synthesis in the baseline approach degenerates into enumerative search.

 \autoref{fig:eval-nodecompose} compares the performance of \toolname against this baseline without decomposition. Here, the $x$-axis shows the time budget per benchmark, and the $y$-axis shows the percentage of benchmarks that can be solved within the given budget. Furthermore, the solid green line corresponds to \toolname, and the dashed blue line corresponds to the baseline without decomposition. As we can see from this figure, having an intermediate specification greatly benefits our synthesis algorithm. In particular, without decomposition, the percentage of benchmarks solved within a 600s (resp. 120s) time-limit drops from 84\% (resp. 80\%) to 60\% (resp. 49\%).

\subsection{Evaluating Table Transformation Algorithm}

In this section, we evaluate the impact of using our new table transformation algorithm over an existing technique that addresses the same problem. To perform this evaluation, we use a variant of \toolname that we refer to \toolvar\ that uses  {\sc Morpheus}~\cite{morpheus} as its table transformation back-end. However, since the original {\sc Morpheus} tool is written in C++, we instead use a newer implementation of {\sc Morpheus} written in Python~\cite{trinity} (by the original {\sc Morpheus} authors). Furthermore, since {\sc Morpheus} does not support our table inclusion constraints, we "translate" the generated intermediate specification to {\sc Morpheus'} constraint language. In particular, given an intermediate specification $\phi$, our "translation" infers the strongest formula expressible in {\sc Morpheus}' language, which consists of equality and inequality constraints on the number of rows or columns of the output table.

The results of this comparison are presented in \autoref{fig:eval-morpheus}, which plots the number of benchmarks that can be solved within a given time budget for both \toolname and \toolvar. As we can see from this figure, the table transformation synthesizer proposed in this paper yields much better results compared to {\sc Morpheus}. In particular, within a 600s (resp. 120s) time-limit, \toolvar\ can solve 69\% (resp. 58\%) of the benchmarks compared to 84\% (resp. 80\%) for \toolname.

\subsection{Example Tasks}

To give the reader some intuition about the class of  tasks that can be automated using \toolname, we highlight three representative visualization tasks from our benchmark set.

\begin{figure}[t]
    \centering
    \includegraphics[width=1\linewidth]{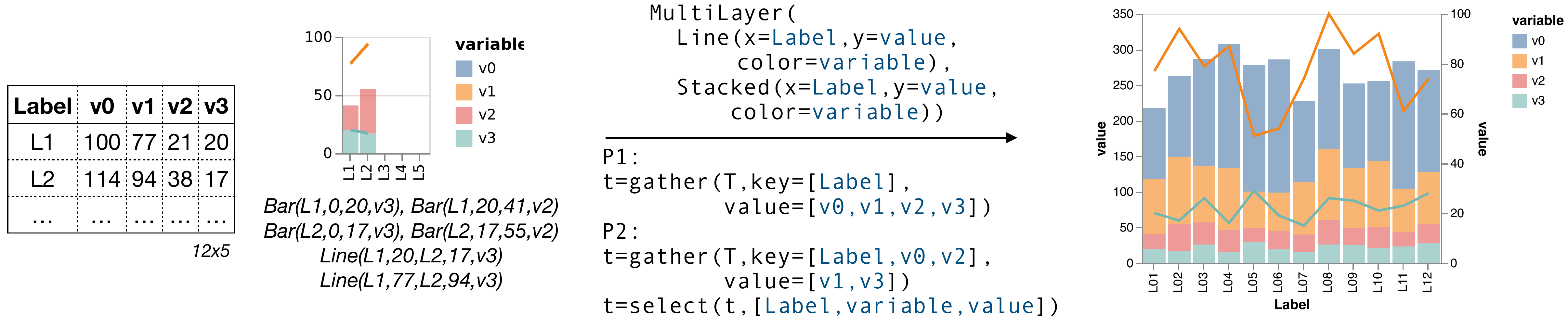}
    \caption{Illustration of visualization task \#1.}
    \label{fig:example-task-1}
\end{figure}

\paragraph{Task \#1} \autoref{fig:example-task-1} shows a visualization task involving multiple layers, consisting of a stacked bar chart and a (multi-layered) line chart. The left-hand side of the figure shows the original data source, and right next to it, we show the visual sketch (and its corresponding visual trace) that we use to automate this visualization task. The synthesized visualization script is indicated on both sides of the arrow (visual program on top; table transformation at the bottom). Finally, the right-most part of the figure shows the resulting visualization that is obtained by applying the synthesized script to the input table. 

Observe that the synthesized visual program refers to columns such as \code{variable} and \code{value} that do not exist in the original table and that are introduced by the table transformation program. Further, as discussed in \autoref{sec:impl}, \toolname synthesizes as many table transformation programs as there are layers; thus, we have two separate table transformation programs for this example. The visualization shown on the right is one of the top-$2$ visualizations produced by \toolname for this example.

\begin{figure}[t]
    \centering
    \includegraphics[width=1\linewidth]{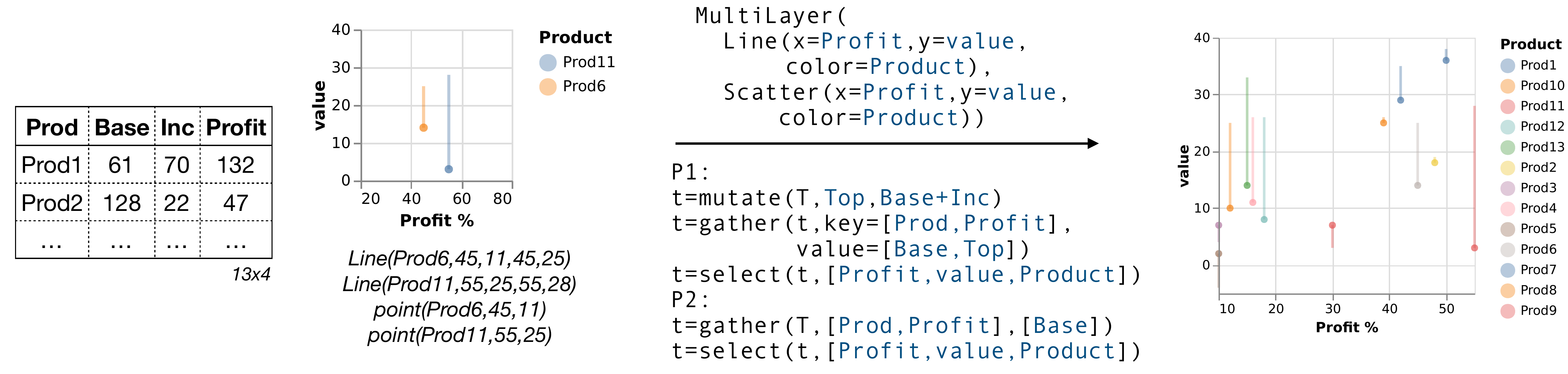}
    \caption{Illustration of task \#2.}
    \label{fig:example-task-2}
\end{figure}

\paragraph{Task \#2} \autoref{fig:example-task-2} shows a scenario in which a user has data on corporate profits and wants to generate a so-called "cherry chart". Since most visualization libraries do not have a "cherry chart" primitive, generating this plot requires layering a line chart with a scatter plot. The left half of \autoref{fig:example-task-2} shows the input to \toolname, and right half shows the synthesized programs and the corresponding visualization of the entire dataset. As in the previous example, we have multiple table transformation programs, one for each layer, and both the visual program and the table transformation programs refer to a column called \code{value} that does not exist in the original table and that is introduced by the \code{gather} operation. In this case, the intended visualization shown on the right is top result returned by \toolname.

\begin{figure}[t]
    \centering
    \includegraphics[width=1\linewidth]{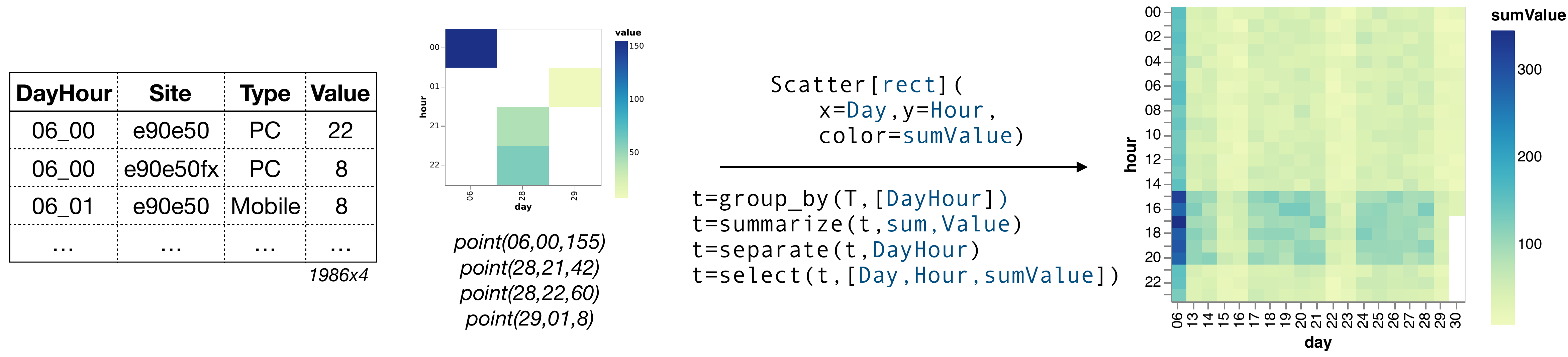}
    \caption{Illustration of task \#3.}
    \label{fig:example-task-3}
\end{figure}

\paragraph{Task \#3} \autoref{fig:example-task-3} shows a visualization task that requires drawing a so-called "heat map" that visualizes the number of visits to various websites at each hour  on different days. In this case, the input data is very large and contains close to $2000$ rows. Furthermore, the corresponding data transformation program is quite complex and requires both computation as well as reshaping. Specifically, the table transformation program computes the total number of website visits for each one-hour time period, which is stored in a column called \code{sumValue} introduced by \code{summarize}. The visual program refers to this new  \code{sumValue} column to generate the desired heat map. Given the visual trace shown in \autoref{fig:example-task-3}, the intended visualization (on the right) is ranked within the top $20$ visualizations, but if we provide a visual sketch with $8$ elements instead of $4$, then the intended visualization is ranked number one.

\subsection{Discussion of Limitations}
\label{sec:limitations}

As reported in \autoref{sec:key-results}, \toolname cannot synthesize 13 of the 83 benchmarks within a time limit of $10$ minutes. To understand the limitations of \toolname in practice, we manually inspected these benchmarks and explain the insights we gained from our examination. Specifically, we highlight two reasons that are responsible for \toolname not finding the desired  visualization within the given time budget. 

\begin{itemize}
\item \emph{Size of the input table.} The size of the input table can affect the performance of \toolname in two ways. First, the search space covered by the table transformation language grows exponentially as we increase the size of the table. This is because constructs in the table transformation language use names of columns as arguments, and, furthermore, rows can become columns during the reshaping process. Second, the table transformation synthesis engine tracks table inclusion constraints where one size of the inclusion is a concrete table. Thus, the larger the initial input table, the more overhead associated with program analysis.


\item \emph{Complex table transformations.} Some tasks in our benchmark suite require very sophisticated table transformations that \toolname is unable to explore within the given time budget. For instance, some benchmarks that \toolname cannot solve within 10 minutes require a combination of relational operations, string manipulation, column-wise arithmetic computations, and table pivoting.
\end{itemize}
\medskip

In addition, recall from \autoref{sec:benchmarks} that approximately 20\% of the visualization tasks  we inspected are not expressible in our visualization language. These benchmarks fall into roughly three classes: (1) visualizations involving continuous functions, (2) visualizations that require custom shapes provided by the user (e.g., emoji icons), and (3) visualizations that cannot be placed in a standard coordinate system, (e.g., tree-maps and parallel coordinates).


\section{Related Work}

In this section, we survey closely related work on data visualization and program synthesis.

\paragraph{Automation for visualization} There has been significant recent interest in (semi-)automating various types of visualization tasks. These efforts include both visualization recommendation systems as well as visualization exploration tools.
Among these, visualization recommendation systems like Draco~\cite{DBLP:journals/tvcg/MoritzWNLSHH19}, CompassQL~\cite{DBLP:conf/sigmod/WongsuphasawatM16}, and ShowMe~\cite{DBLP:journals/tvcg/MackinlayHS07}   recommend top completions of an incomplete visualization program. On the other hand, visualization exploration tools, such as VisExamplar~\cite{DBLP:journals/tvcg/SaketKBE17}, Visualization-by-Sketching~\cite{DBLP:journals/tvcg/SchroederK16}, Polaris~\cite{DBLP:journals/cacm/StolteTH08}, and Voyager~\cite{DBLP:journals/tvcg/WongsuphasawatM16,DBLP:conf/chi/WongsuphasawatQ17}, aim to generate diverse visualizations based on user demonstrations, which can include  graphical sketches, manipulation trajectories, and constraints. 
 All of these existing systems focus on creating visualizations for a fixed dataset and  require the user to prepare the data for a specific visualization API. In contrast, our approach also handles the data preparation and wrangling aspect of data visualization and can be viewed as being  more user-friendly in this respect. However, it is worth noting that many of these systems are complementary to the approach proposed in this paper. For example,  our approach can be used in conjunction with existing systems to rank synthesis results that are consistent with the demonstration. Furthermore, our approach can work with existing visualization demonstration interfaces~\cite{DBLP:journals/cgf/SatyanarayanH14,DBLP:journals/tvcg/SaketKBE17} to reduce user effort in creating a visual sketch.

\paragraph{Automating table transformations} Our technique for synthesizing table transformations is related to several recent techniques for 
automating data wrangling~\cite{harris2011spreadsheet,scythe,morpheus,DBLP:conf/pldi/FengMBD18,sql2,DBLP:conf/sigmod/TranCP09}. 
Among these, Scythe generates SQL queries from input-output examples and prunes the search space by grouping partial queries into equivalence classes~\cite{scythe}. The Morpheus system automates table transformation tasks that arise in R programming and leverages logical  specifications of R library functions to prune the search space using SMT-based reasoning~\cite{morpheus}. Morpheus's successor, {\sc Neo}, generalizes this technique to other domains and further uses logical specifications to learn from failed synthesis attempts~\cite{DBLP:conf/pldi/FengMBD18}. A unifying theme among all these prior efforts is that the specification is a pair of concrete input and output tables. In contrast, our specification does not involve a concrete output table but rather a set of table inclusion constraints; furthermore, our approach works with the original (potentially very large) dataset and does not require the user to craft a small representative input table.  The large input table assumption is likely to be problematic for systems like Scythe and Morpheus that require  evaluating the partial program on the input table. Furthermore, since the output specification is much weaker in our context compared to existing systems, forward reasoning alone is not sufficient to meaningfully prune the search space, as demonstrated in our evaluation.

\paragraph{Program analysis for program synthesis.} Given the large search space that must be explored by program synthesizers, a common trick is to perform lightweight program analysis to prune the search space~\cite{blaze,synquid,regex,l2}. The particular flavor of program analysis varies between different synthesizers and ranges from domain-specific deduction~\cite{l2,regex} to
abstract interpretation~\cite{blaze} to SMT-based reasoning~\cite{DBLP:conf/pldi/FengMBD18,synquid}. Furthermore, some of these techniques leverage program analysis to construct a compact version space~\cite{blaze,flashmeta} while others use it to prune partial programs in enumerative search~\cite{DBLP:conf/pldi/FengMBD18,regex}. Similar to these efforts, we also use program analysis to prove infeasibility of partial programs but with two key differences: First, our analysis is tailored to table transformation programs and infers inclusion constraints between tables. Second, since neither forward nor backward reasoning is sufficient to meaningfully prune the search space on their own, we use bi-directional analysis to improve pruning power without having to resort to heavy-weight semantics motivated by prior work in program analysis~\cite{DBLP:conf/popl/RepsHS95,DBLP:conf/pldi/ChandraFS09,DBLP:conf/sas/DhurjatiDY06} and verification~\cite{DBLP:journals/pacmpl/WangCB18}. In this respect, our synthesis method is similar to {\sc Synquid}~\cite{synquid} which uses a form of bidirectional refinement type checking to prune its search space. However, unlike {\sc Synquid} which requires precise refinement type specifications of components, our method uses lightweight semantics that are tailored specifically for our table transformation DSL. Furthermore, in contrast to {\sc Synquid} which leverages an SMT solver, our method uses a custom, and deliberatively incomplete, solver for checking satisfiability at low cost.

\paragraph{Compositional program synthesis.} As mentioned throughout the paper, our technique decomposes the synthesis task into two separate sub-problems. In this respect, our method is similar to prior efforts on compositional program synthesis~\cite{l2,synquid,compNL,flashmeta,DBLP:conf/asplos/PhothilimthanaT16,DBLP:journals/pacmpl/MainaMFPWZ18,DBLP:journals/pacmpl/MiltnerFPWZ18}. Among these techniques, $\lambda^2$ uses domain knowledge about the DSL constructs to infer input-output examples for sub-expressions whenever feasible~\cite{l2}, FlashMeta (and its variants) use inverse semantics of DSL constructs to propagate examples backwards~\cite{flashmeta}, and Optician~\cite{DBLP:journals/pacmpl/MainaMFPWZ18,DBLP:journals/pacmpl/MiltnerFPWZ18} decomposes the synthesis process using DNF regular expression outlines. {\sc Synquid} also tries to decompose the overall specification into sub-goals using a technique referred to as "round-trip type checking"~\cite{synquid}. On a slightly different note, the technique of Raza et al.~\cite{compNL} also performs  synthesis in a compositional way, but it leverages natural language to identify sub-problems and asks the user to provide input-output examples for each auxiliary task. In contrast to all of these techniques, our method decomposes the visualization synthesis task into two sub-problems over \emph{different} DSLs and uses  the inverse semantics of the visualization DSL to infer precise constraints on the input table. The inferred specification is precise in the sense that any table transformation program that satisfies this specification is guaranteed to be a valid solution.

\section{Conclusion}

In this paper, we introduced \emph{visualization-by-example}, a new program synthesis technique for generating visualizations from visual sketches. Given the original raw data and a visual sketch consisting of a few visual elements, our technique can automatically synthesize visualization scripts that yield a visualization  consistent with the user's visual sketch. Our technique decomposes the synthesis problem into two sub-tasks by inferring an intermediate specification in the form of table inclusion constraints. This intermediate specification is then used to guide the synthesis of table transformation programs using a combination of bi-directional program analysis and lightweight inference over table inclusion constraints.

We have implemented the proposed method as a new tool called \toolname that allows users to explore different visualizations for the entire data set based on a small visual sketch. Notably, and unlike any other visualization tool, \toolname can perform any necessary data wrangling tasks, including reshaping and aggregation. We have evaluated \toolname on a benchmark suite consisting of $83$ visualization tasks obtained from on-line forums and tutorials. Given a visual sketch consisting of four visual elements and a time budget of 600 seconds, \toolname can solve 84\% of these tasks. Furthermore, among the 70 tasks that can be solved within the time budget, the desired visualization is ranked within top $5$ in  $76\% $ of the cases. Beyond showing that \toolname can help automate real-world data visualization tasks, our evaluation also confirms the importance of decomposing the synthesis task as well as the necessity of our proposed table transformation synthesizer.

In the near future, we are interested in integrating \toolname with visual demonstration interfaces proposed in the visualization literature. Such interfaces can  make \toolname more user-friendly by providing a graphical user interface that allows users to draw visual sketches rather than write visual traces in a semi-formal language. We also plan to improve the search heuristics underlying \toolname so that visualization scripts that generate the intended visualization are more likely to be explored first.

\appendix

\section{Appendix}

\subsection{Full Visualization Language}

In this section, we present definitions of the full visualization language $\lang_\viz$ and the full trace language $\lang_\trace$ we use in practice for the visualization by example task.

\begin{figure}[ht]
\[
\begin{array}{rlll}
\pvis & = &  \mathsf{MultiPlot}(SP, c_\mathsf{row}, c_\mathsf{col}) ~|~ SP & \text{(Faceted Chart)}\\
\mathit{SP} & = & \mathsf{MultiLayer}(\bar{L}) ~|~ L & \text{(Layered Chart)}\\
\mathit{L} & = & \mathsf{Scatter}[\mathit{mark}](c_x, c_y, c_\mathit{shape}, c_\mathit{color}, c_\mathit{size}, c_\mathit{text}) & \text{(Scatter Plot)}\\
  & | & \mathsf{Line}(c_x, c_y, c_\mathit{width}, c_\mathit{order}, c_\mathit{color}, c_\mathit{detail}) & \text{(Line Chart)}\\
  & | & \mathsf{Bar}(c_x, c_{x_2}, c_y, c_{y_2}, c_\mathit{color}, c_\mathit{width}) & \text{(Bar Chart)}\\
  & | & \mathsf{StackedBar}[\mathit{orient}](c_x, c_h, c_\mathit{color}, c_\mathit{width}) &  \text{(Stacked Bar Chart)} \\
  & | & \mathsf{Area}(c_x, c_{x_2}, c_y, c_{y_2}, c_\mathit{color}) & \text{(Area Chart)}\\
  & | & \mathsf{StackedArea}[\mathit{orient}](c_x, c_h, c_\mathit{color}) &  \text{(Stacked Area Chart)} \\
c & = & \mathit{column} ~|~ \epsilon \\
\mathit{mark} & = & \mathsf{point}~|~\mathsf{circle}~|~\mathsf{text}~|~\mathsf{rect}~|~\mathsf{tick} \\
\mathit{orient} & = & \mathsf{horizontal} ~|~ \mathsf{vertical}
\end{array}
\]
\vspace{-10pt}
\caption{The full visualization language $\lang_\viz$.}
\label{fig:full-vis-lang}
\end{figure}

\autoref{fig:full-vis-lang} defines our full visualization language $\lang_\viz$. A visual program $\pvis$ either creates a grid of multiple plots using the \textsf{MultiPlot} construct or a single plot $SP$ (where grid index for each subplot is determined by its value in column $c_\mathsf{col}$ and $c_\mathsf{row}$). Each plot can in turn consist of multiple layers (indicated by the \textsf{MultiLayer} construct) or a single layer. Each layer is either a scatter plot (\textsf{Scatter[\textit{mark}]}, where the paramter \textit{mark} decides the shape of scatter points), a line chart (\textsf{Line}), a bar chart (\textsf{Bar}), a stacked bar chart (\textsf{StackedBar}), an area chart (\textsf{Area}) or a stacked area chart (\textsf{StackedArea}).  The \textsf{MultiLayer} construct in this language is used to compose \emph{different} kinds of charts in the same plot (e.g., a scatter plot and a line chart), as our visualization language is already rich enough to allow layering the same type of chart within a plot.  

\begin{figure}[ht]
\[
\begin{array}{rllll}
\mathit{\trace} & = & \{\mathit{e}_1, \dots, \mathit{e}_n\}\\
\mathit{e} & = 
       & \mathsf{barV}(a_{x}, a_{y_1}, a_{y_2}, a_\mathit{width}, a_\mathit{color}, a_\mathit{col}, a_\mathit{row}) & \text{(Vertical Bar)}\\
      &|& \mathsf{barH}(a_{y}, a_{x_1}, a_{x_2}, a_\mathit{width}, a_\mathit{color}, a_\mathit{col}, a_\mathit{row}) & \text{(Horizontal Bar)}\\
      &|& \mathsf{point}(a_{x}, a_{y},a_\mathit{shape}, a_\mathit{color}, a_\mathit{size}, a_\mathit{col}, a_\mathit{row}) \\
      &|& \mathsf{line}(a_{x_1}, a_{y_1}, a_{x_2}, a_{y_2}, a_\mathit{width}, a_\mathit{color}, a_\mathit{col}, a_\mathit{row})\\
      &|& \mathsf{area}(a_{x_{tl}}, a_{y_{tl}}, a_{x_{bl}}, a_{y_{bl}}, a_{x_{tr}}, a_{y_{tr}}, a_{x_{br}}, a_{y_{br}}, a_\mathit{color}, a_\mathit{col}, a_\mathit{row})\\
\mathit{mark} & = & \mathsf{point}~|~\mathsf{circle}~|~\mathsf{text}~|~\mathsf{rect}~|~\mathsf{tick} \\
\end{array}
\]
\vspace{-10pt}
\caption{The full trace language $\lang_\trace$, where metavariable $a$ refers to constants.}
\label{fig:full-trace-lang}
\end{figure}

Visual traces encode semantics of visualizations. A visual trace, denoted $\trace$, is a set of basic visual elements, i.e., $\mathsf{point}$, $\mathsf{line}$, $\mathsf{barH}$ (horizontal bar), $\mathsf{barV}$ (vertical bar), or $\mathsf{area}$, together with the  attributes of each element (position, size, color, etc.). \autoref{fig:full-trace-lang} shows the language in which we express visual traces. Here, $e$ denotes a visual element, and $a$ is an \emph{attribute} of that element:

\begin{itemize}\itemsep-1pt
\item \emph{Color attribute:} This attribute, denoted $a_\emph{color}$ specifies the color of a visual element.
\item \emph{Position attributes:}  Position attributes, such as $a_x, a_{x_1}, a_{y_2}$ etc., specify the canvas positions for a visual element. For \textsf{line}, $(a_{x_1}, a_{y_1})$ and $(a_{x_2}, a_{y_2})$ specifies the starting and the end points of a line segment. For the \textsf{bar} visual element, $a_{y_1}, a_{y_2}$ specify the start and end $y$-coordinates of a (vertical) bar. Similarly, attributes $a_{x_{tl}}, a_{y_{tl}}, a_{x_{bl}}, a_{y_{bl}}, a_{x_{tr}}, a_{y_{tr}}, a_{x_{br}}, a_{y_{br}}$ specity $x,y$ positions of top left / bottom left / top right / bottom right corners of an $\mathsf{area}$ element.
\item  \emph{Size / Shape attributes:} Attributes $a_\emph{size}$ and $a_\emph{shape}$ specify the size and the shape variation of a given \textsf{point} element in a scatter plot.
\item  \emph{Width attribute:} The attribute $a_\emph{width}$ specifies the width of a given \textsf{barH} / \textsf{barV} / \textsf{Line} element.
\item \emph{Subplot attribute:} Attributes $a_\emph{col}, a_\emph{row}$ specify the subplot that a given visual element belongs to. For instance, a point with $a_\emph{col}=1$ and $a_\emph{row}=2$ belongs to the subplot located the first column and second row of visualization containing multiple plots.
\end{itemize}

\begin{acks}
This work is supported in part by the National Science Foundation through grants ACI OAC--1535191, IIS-1546083, IIS-1651489 and IISOAC-1739419; DARPA awards FA8750--14--C--0011, FA8750--16--2--0032 and FA8750-16-2-0032; the Intel and NSF joint research center for Computer Assisted Programming for Heterogeneous Architectures (CAPA NSF CCF-1723352); the CONIX Research Center, one of six centers in JUMP, a Semiconductor Research Corporation (SRC) program sponsored by DARPA CMU 1042741-394324 AM01; DOE award DE-SC0016260; and gifts from Adobe, Mozilla, Nokia, Qualcomm, Google, Huawei, and NVIDIA. We would also like to thank anonymous reviewers for their valuable comments on paper revising.
\end{acks}

\bibliography{main}

\end{document}